\documentclass{aa}  
\usepackage{graphicx}
\usepackage{natbib}
\bibpunct{(}{)}{,}{a}{}{;}
\usepackage{amssymb}
\usepackage{txfonts}

\begin{document}
   \title{Dense and warm molecular gas in the envelopes and outflows of southern low-mass protostars}

   \author{T.A. van Kempen
          \inst{1,2}
	  \and
	  E.F. van Dishoeck\inst{1,3}
          \and
          M.R. Hogerheijde\inst{1}
          \and 
          R. G\"usten\inst{4}}
   \offprints{T. A. van Kempen}

   \institute{$^1$Leiden Observatory, Leiden University, P.O. Box 9513,
              2300 RA Leiden, The Netherlands\\	      
              $^2$ Center for Astrophysics, 60 Garden Street, MS 78, Cambridge, MA 02138, USA      \\
	      $^3$ Max Planck Institut f\"ur Extraterrestrische 
             Physik (MPE), Giessenbachstr.\ 1, 85748 Garching, Germany \\
             $^4$ Max Planck Institut f\"ur Radioastronomie, Auf dem H\"ugel 69, D-53121 Bonn, Germany\\
             \email{tvankempen@cfa.harvard.edu}
                     }

   \date{June 2009}
\titlerunning{Molecular gas associated with southern protostars}




\def\placeFigureChapterFourOneTwo{
\begin{figure*}[!th]
\begin{center}
\includegraphics[width=120pt]{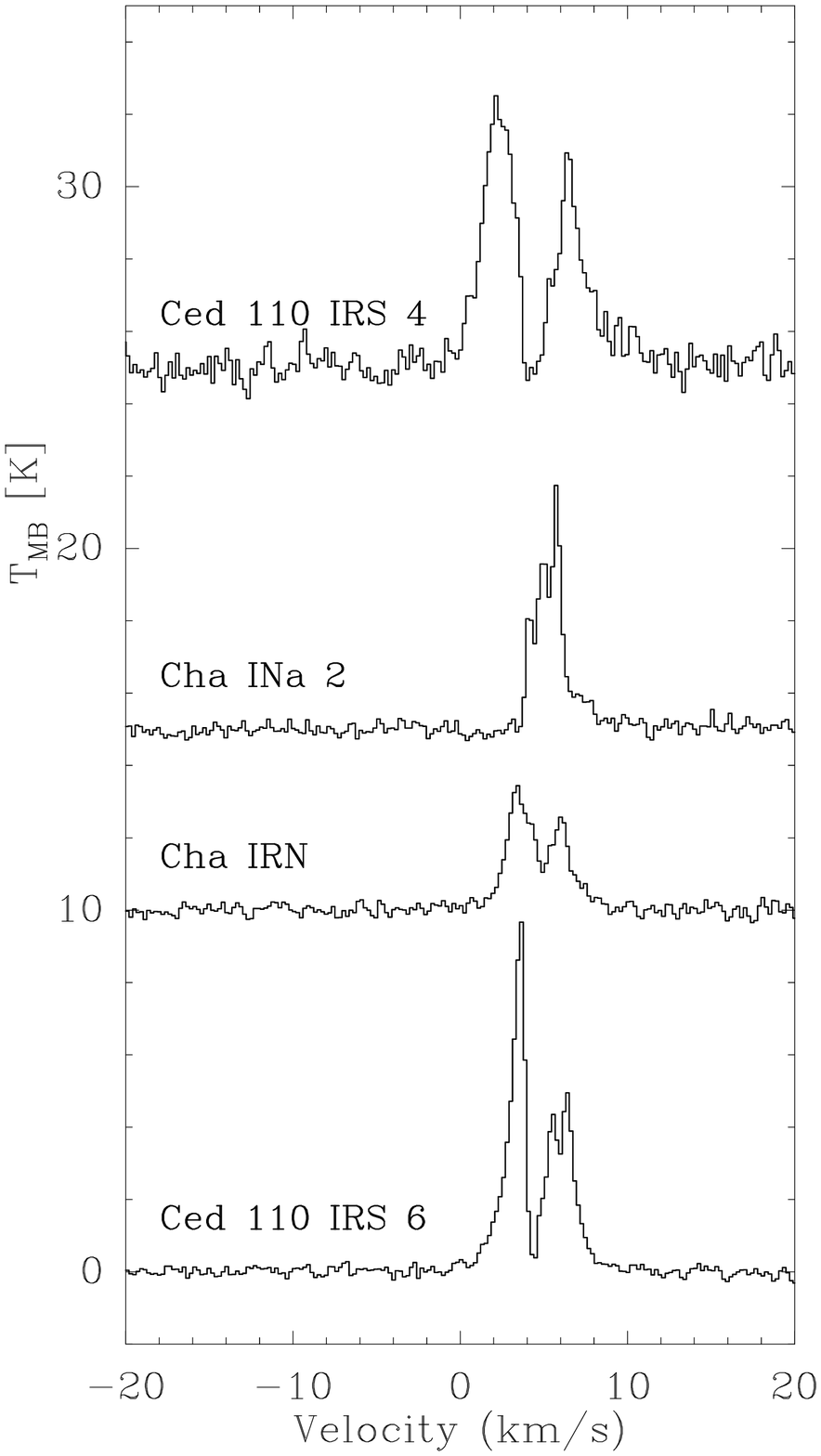}
\includegraphics[width=130pt]{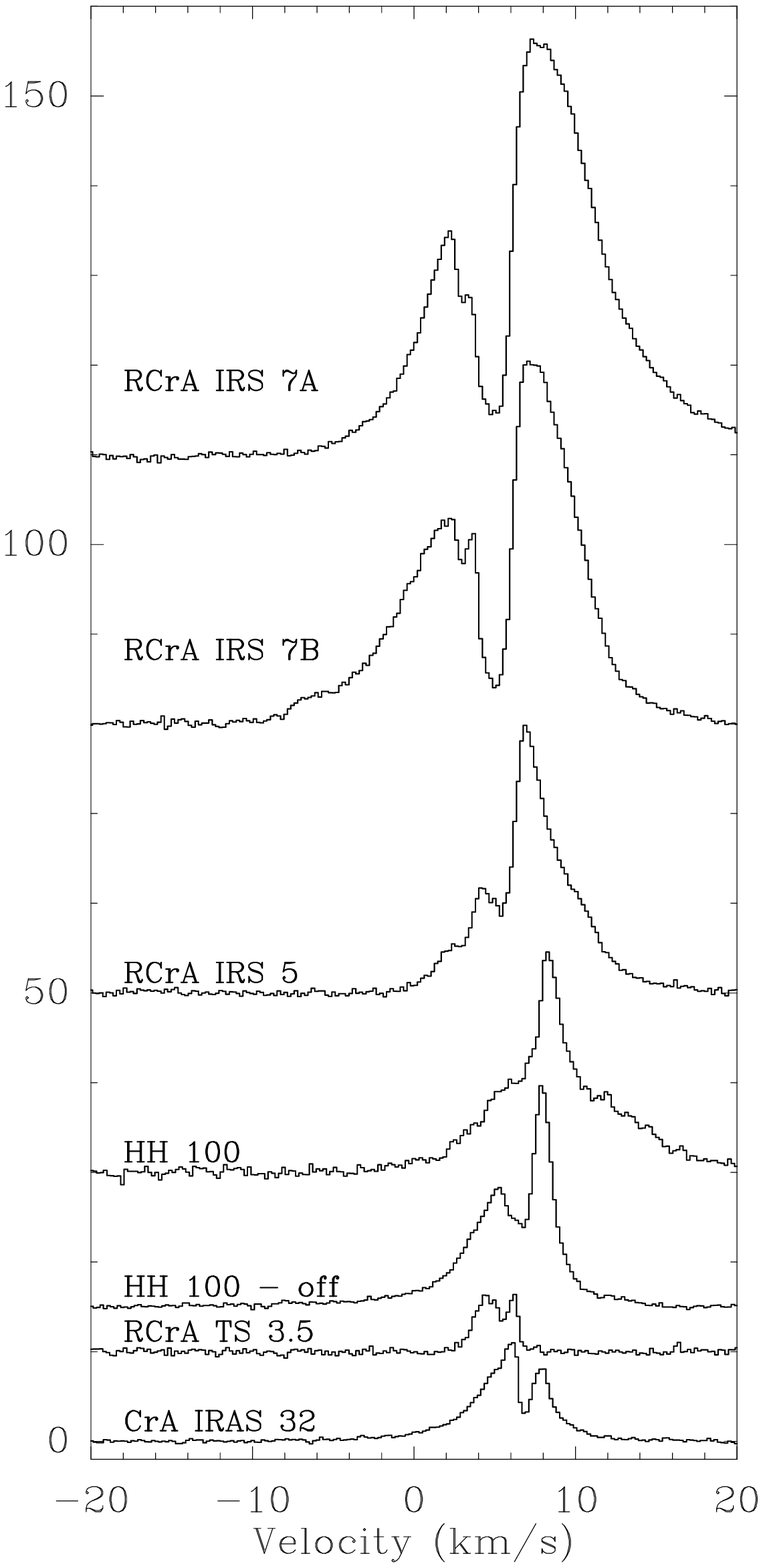}
\includegraphics[width=120pt]{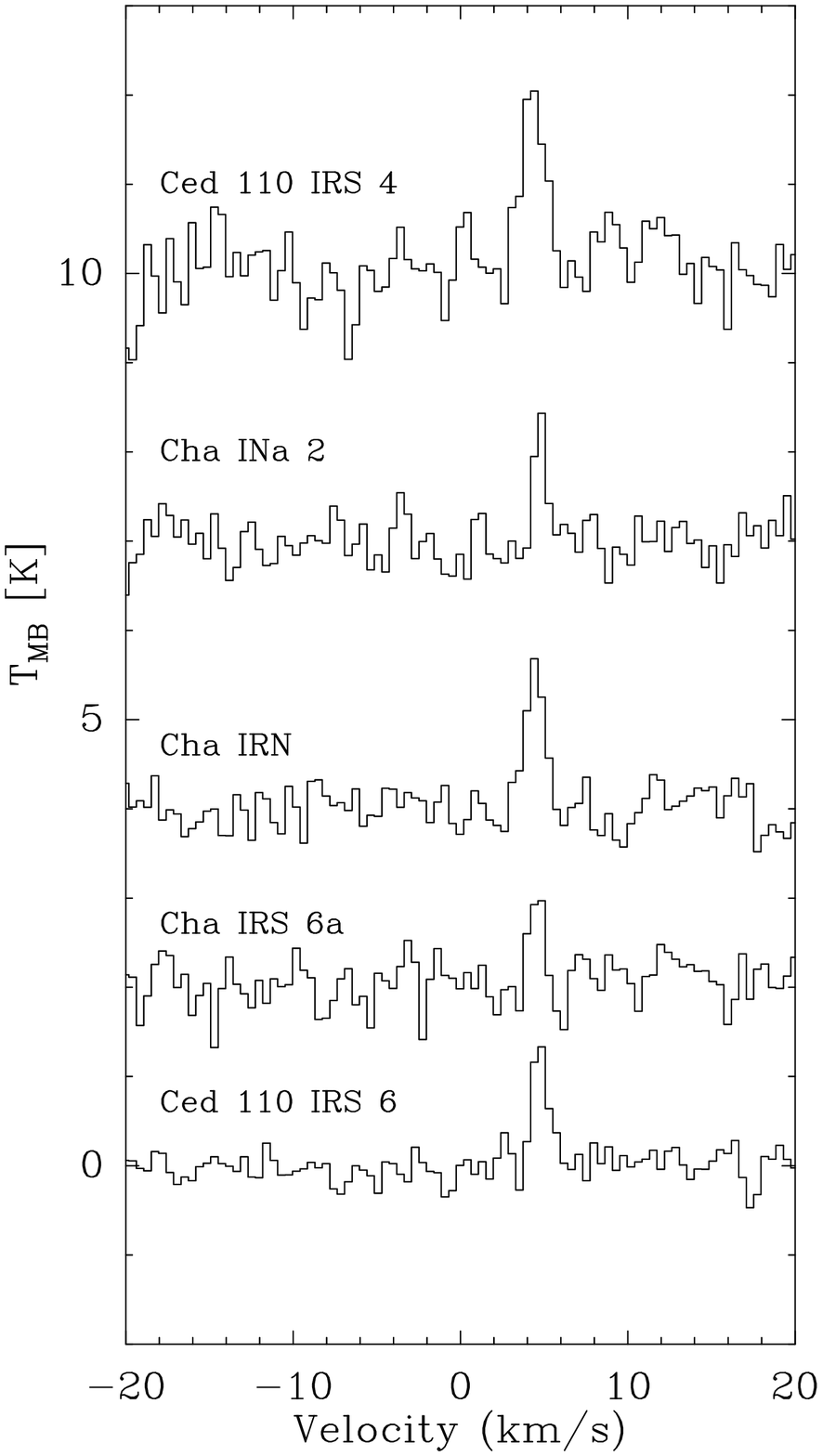}
\includegraphics[width=120pt]{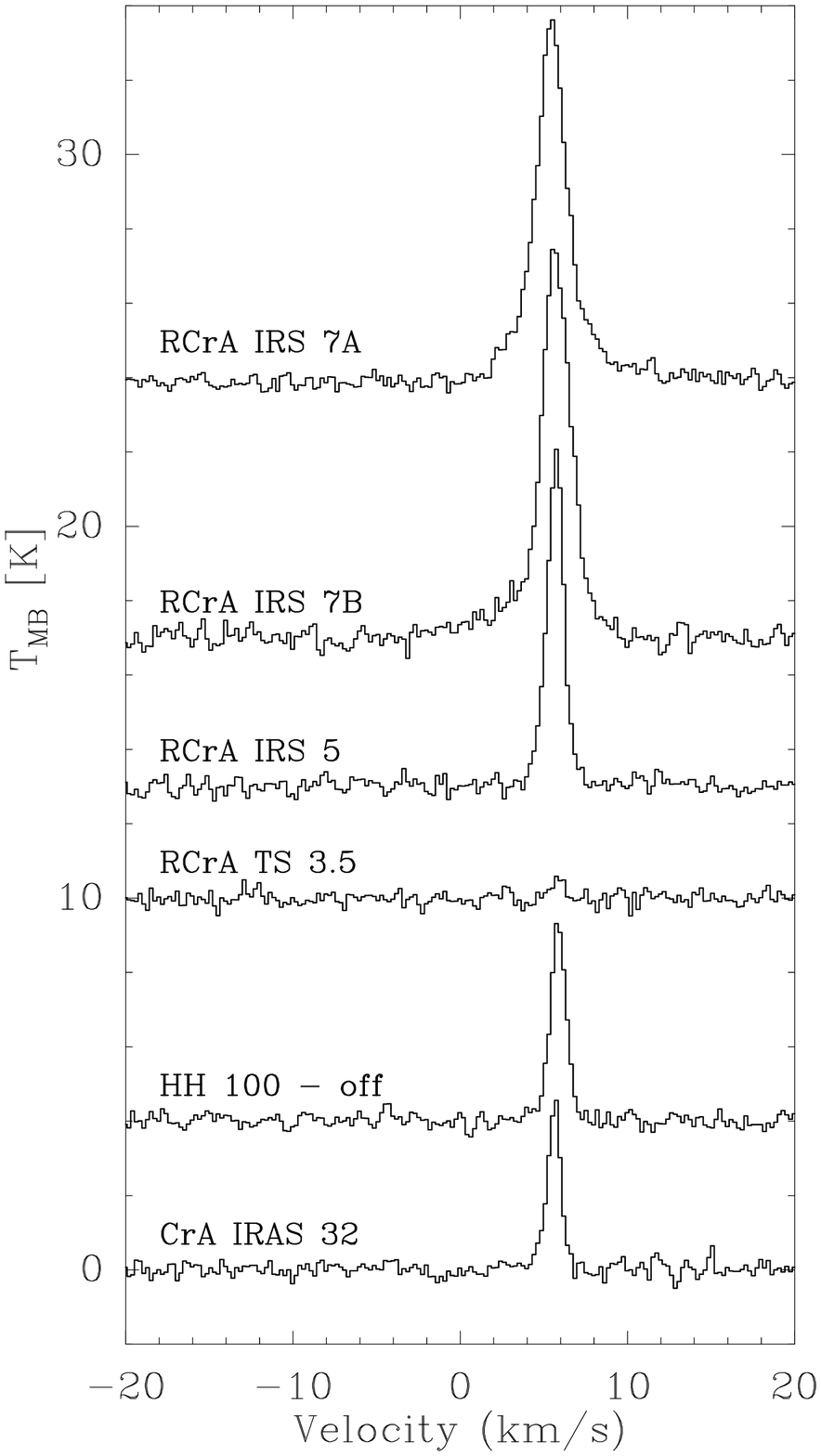}
\end{center}
\caption{$^{12}$CO 3--2 spectra of the sources in Chamaeleon ({\it far left}) and Corona Australis ({\it middle left}) and C$^{18}$O 3--2 spectra of the sources in Chamaeleon ({\it middle right}) and Corona Australis ({\it far right}). Spectra in this and subsequent figures are offset for clarity.}
\label{4:fig:COC18O}
\end{figure*}
}

\def\placeFigureChapterFourThreeFour{
\begin{figure*}[!th]
\begin{center}
\includegraphics[width=120pt]{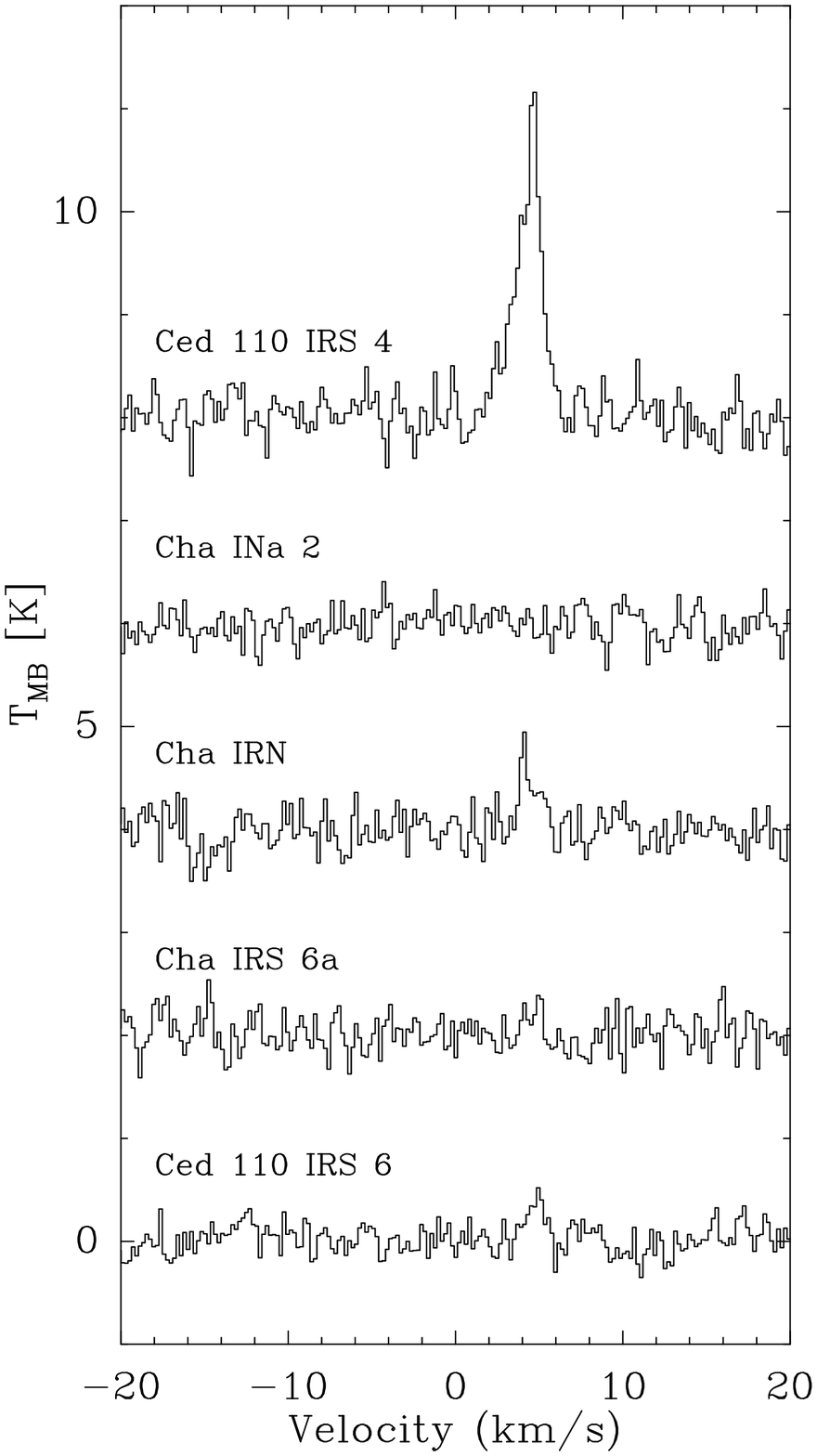}
\includegraphics[width=120pt]{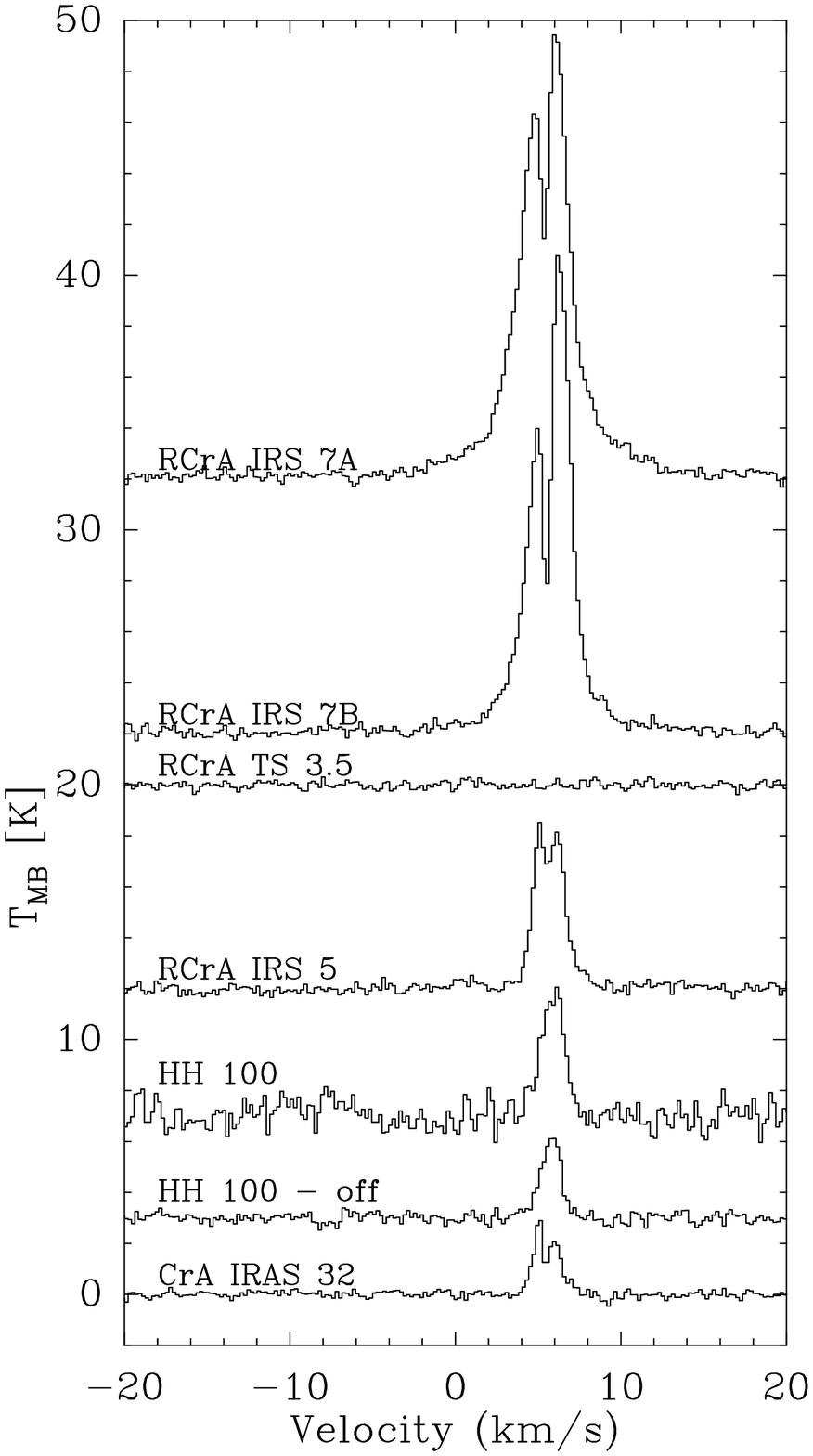}
\includegraphics[width=120pt]{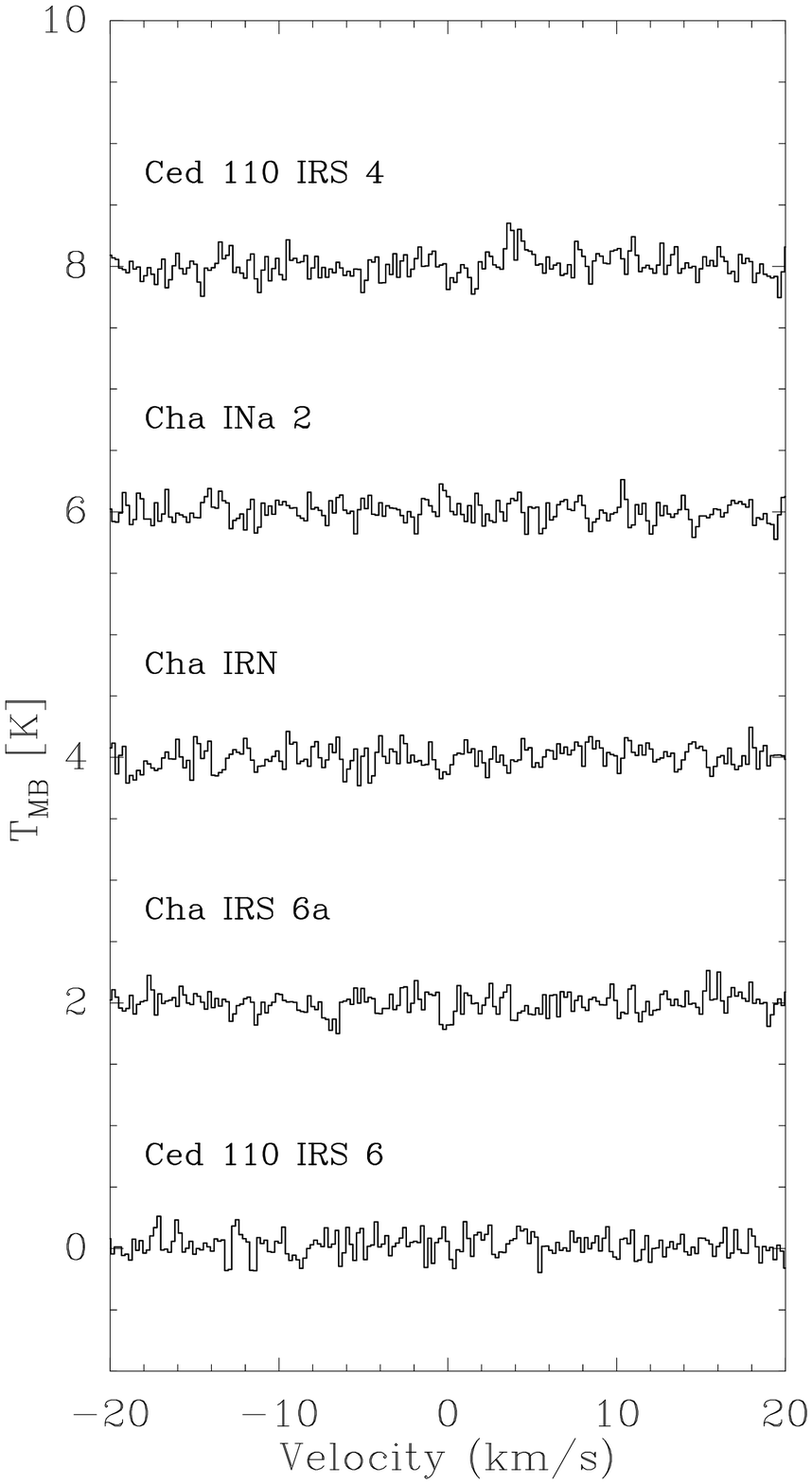}
\includegraphics[width=120pt]{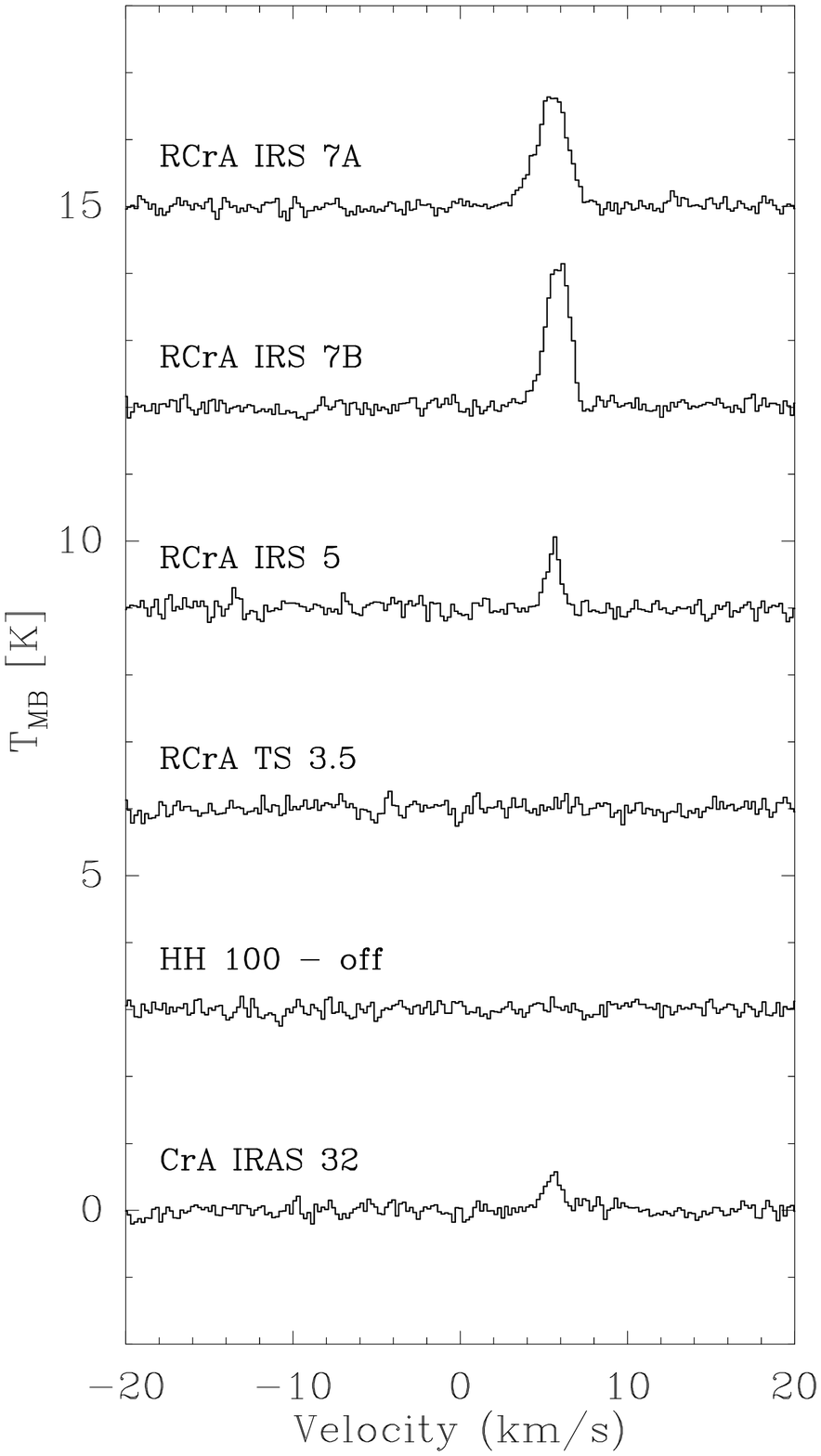}
\end{center}
\caption{HCO$^+$ 4--3 spectra of the sources in Chamaeleon ({\it far left}) and Corona Australis ({\it middle left}), as well as H$^{13}$CO$^+$ 4--3 spectra of the sources in Chamaeleon ({\it middle right}) and Corona Australis ({\it far right}).}
\label{4:fig:HCOH13CO}
\end{figure*}
}

\def\placeFigureChapterFourFive{
\begin{figure}[h]
\begin{center}
\includegraphics[width=120pt]{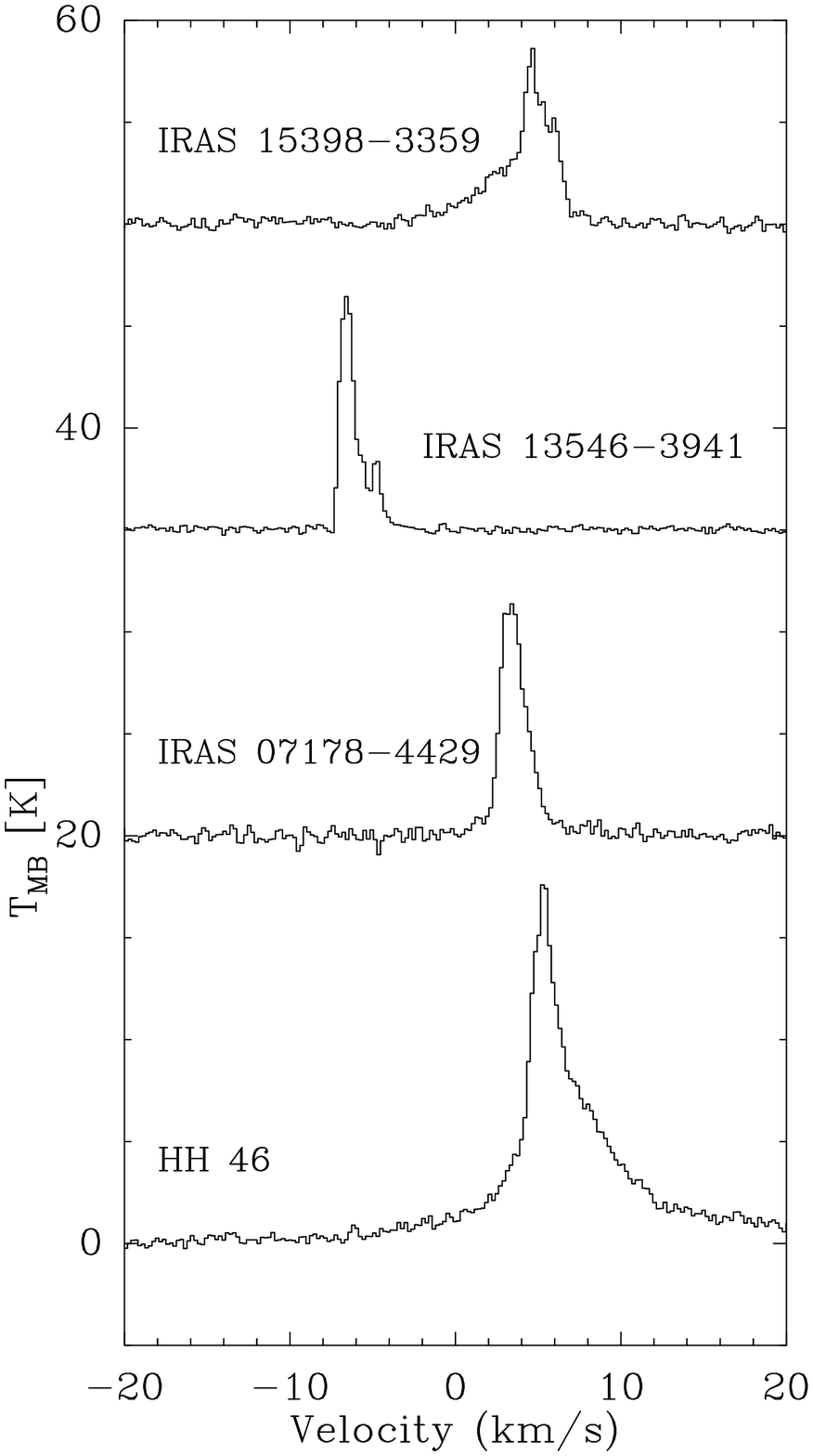}
\includegraphics[width=120pt]{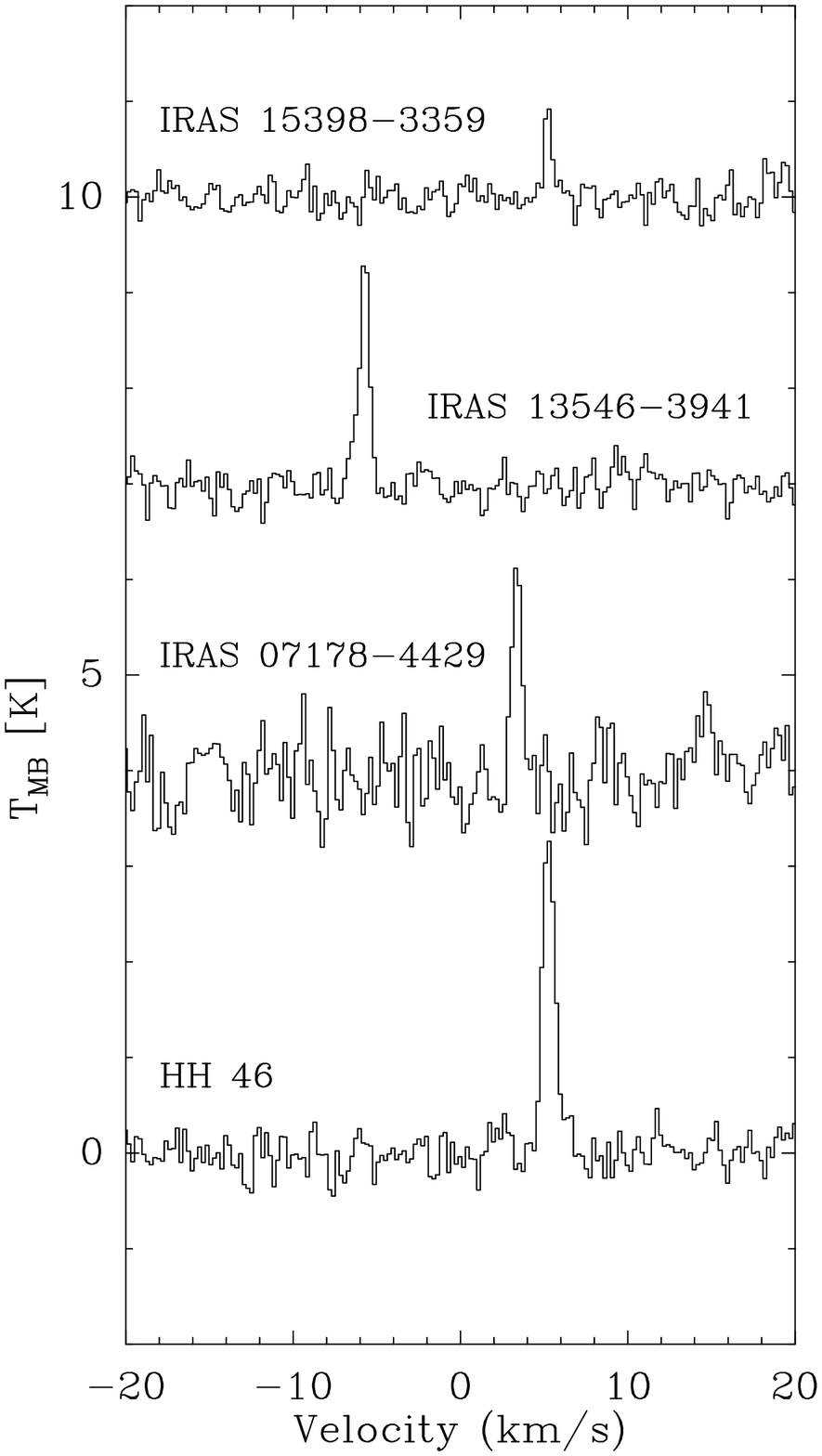}
\end{center}
\caption{$^{12}$CO $J$=3--2 ({\it left}) and C$^{18}$O 3--2 ({\it right}) spectra for the isolated sample. }
\label{4:fig:isoCO}
\end{figure}
}
\def\placeFigureChapterFourSix{
\begin{figure}[h]
\begin{center}
\includegraphics[width=120pt]{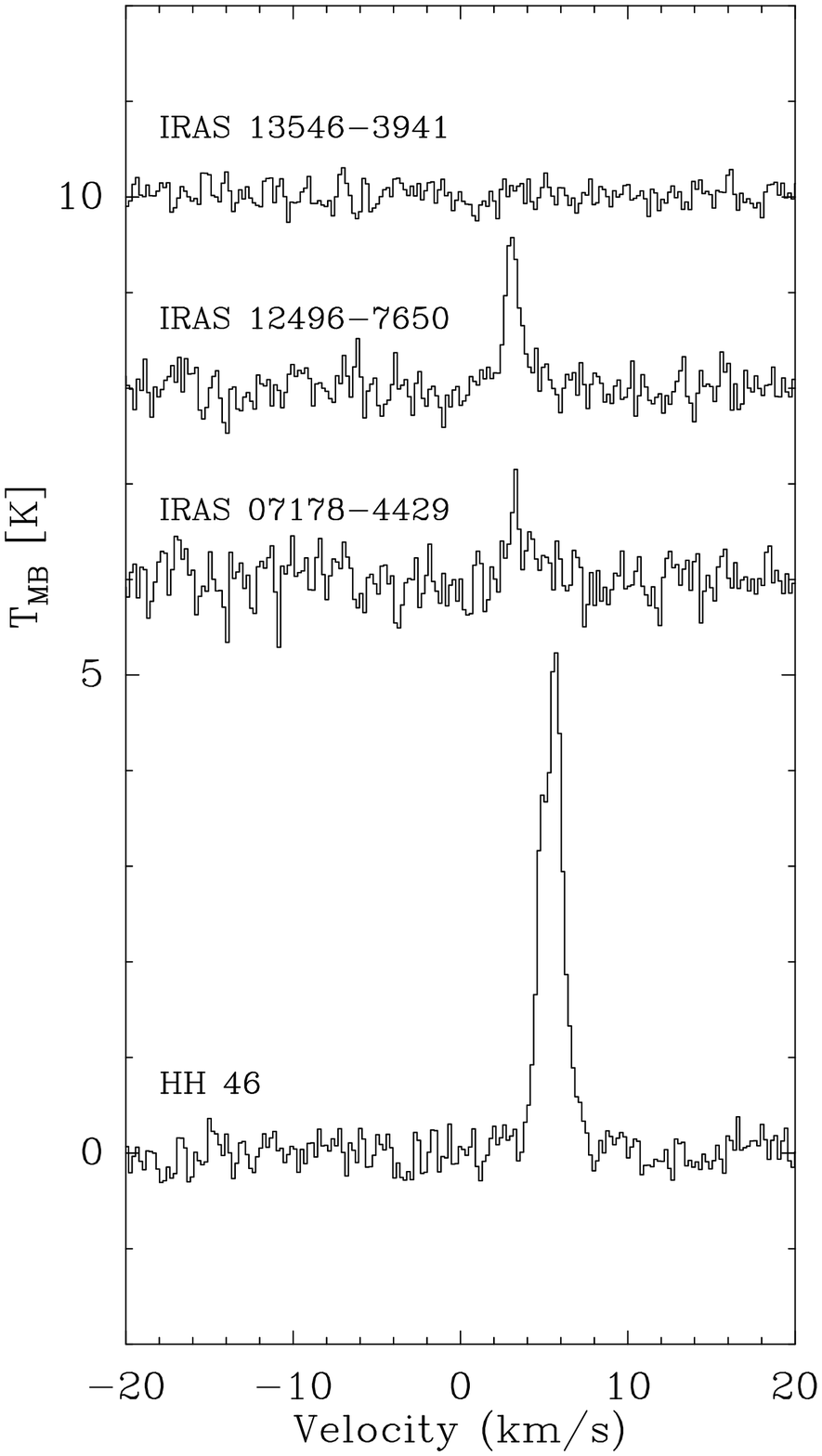}
\includegraphics[width=120pt]{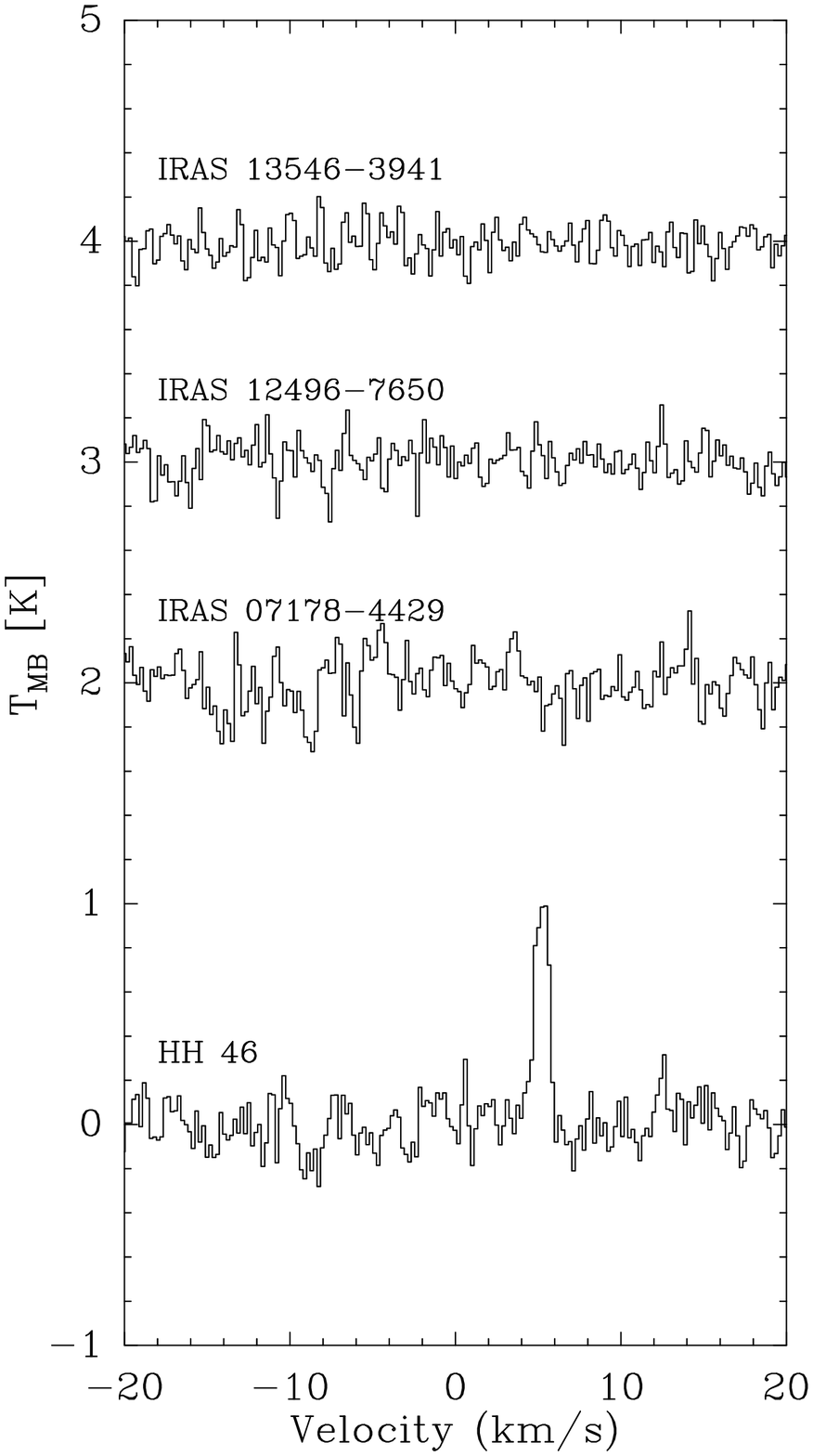}
\end{center}
\caption{HCO$^+$ $J$=4--3 ({\it left}) and H$^{13}$CO$^+$ 4--3 ({\it right}) spectra for the isolated sample. }
\label{4:fig:isoHCO}
\end{figure}
}

\def\placeFigureChapterFourSeven{
\begin{figure}[h]
\begin{center}
\includegraphics[width=83pt]{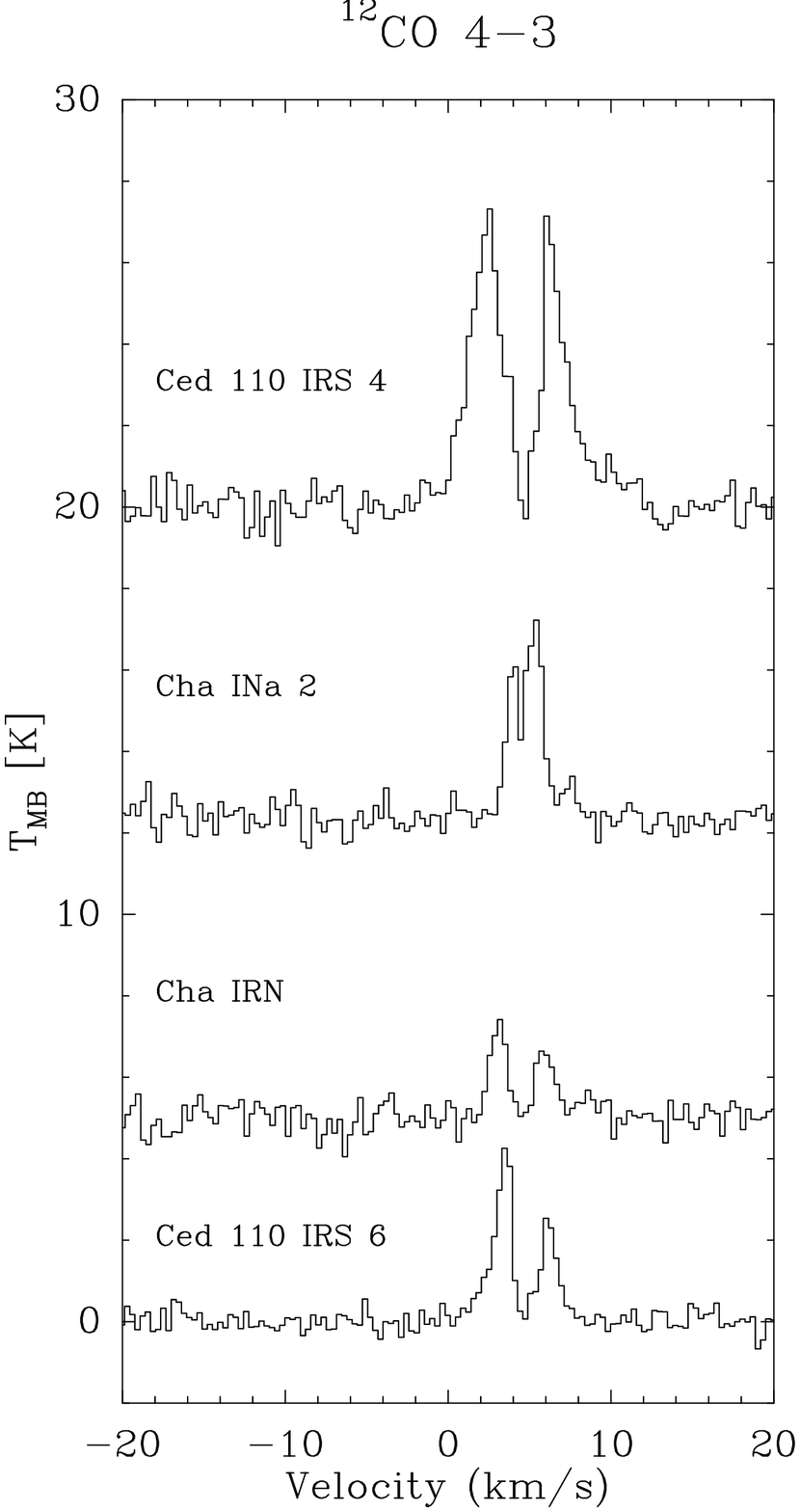}
\includegraphics[width=83pt]{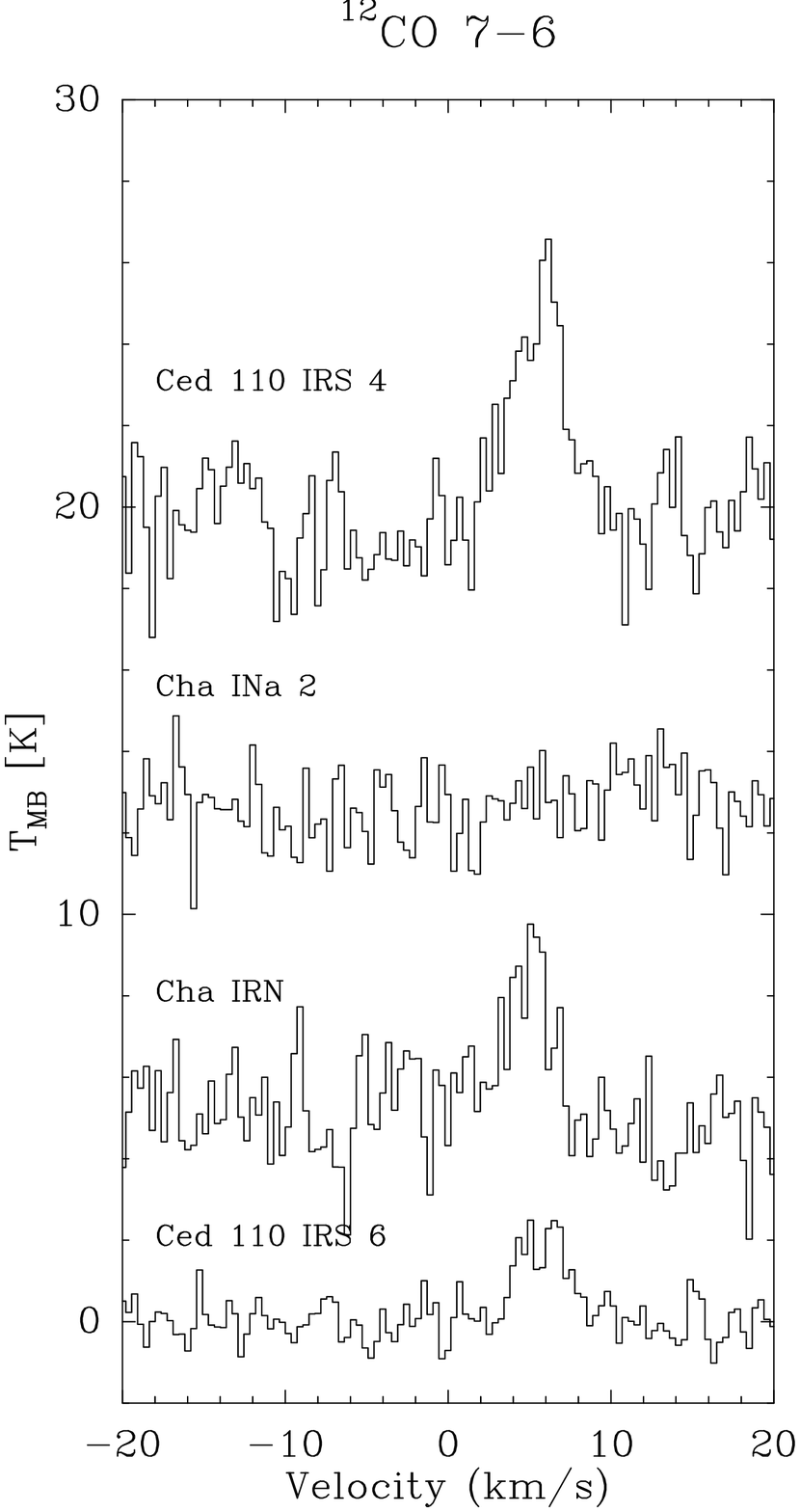}
\includegraphics[width=83pt]{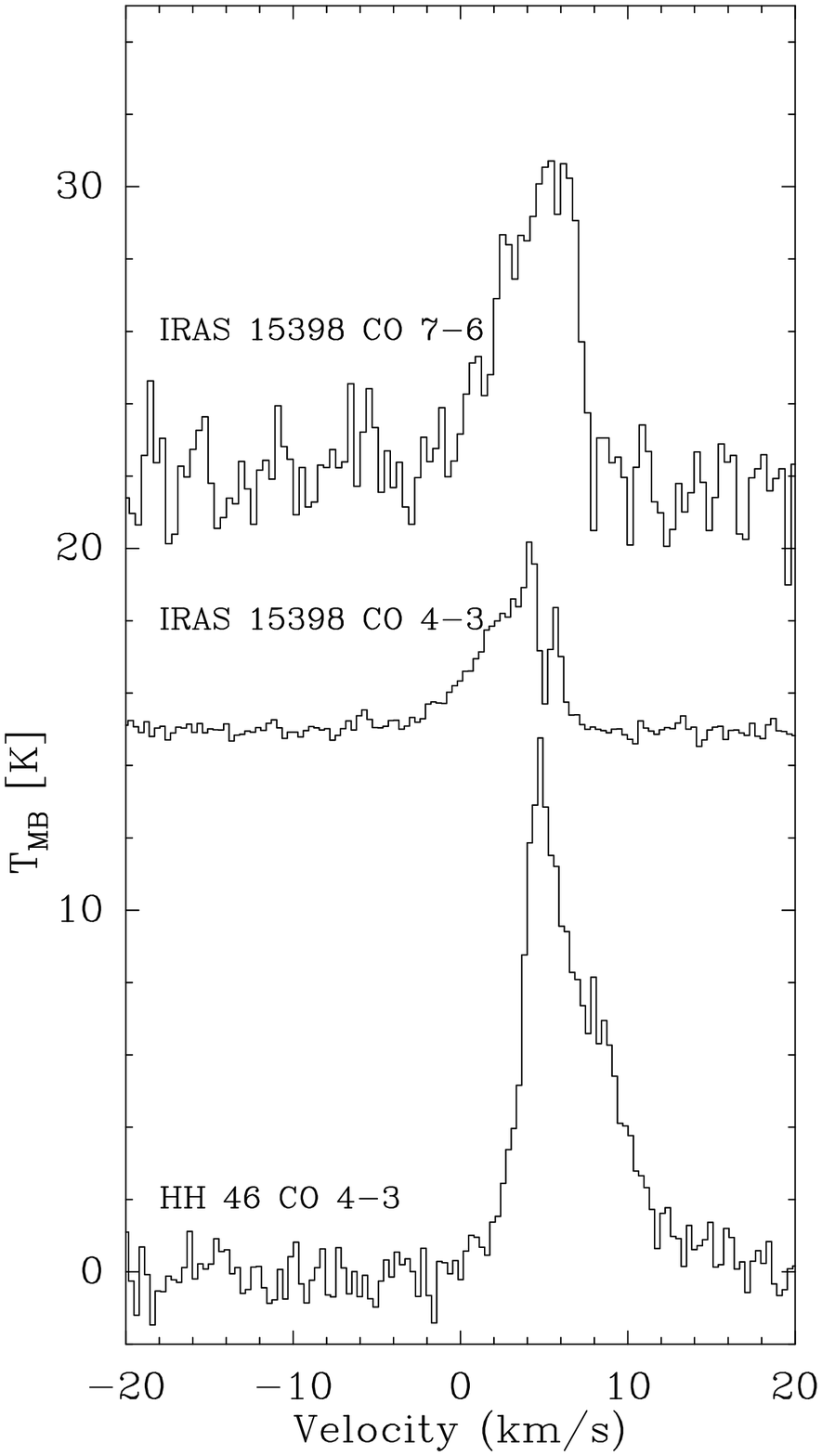}
\end{center}
\caption{$^{12}$CO $J$=4--3 (\textit{left}) and 7--6 (\textit{middle}) spectra of the sources in Chamaeleon, as well as the  spectra of HH 46 and IRAS 15398-3359 taken with FLASH (\textit{right}).}
\label{4:fig:FLASHCO}
\end{figure}
}

\def\placeFigureChapterFourEight{
\begin{figure*}[!htp]
\begin{center}
\includegraphics[width=200pt]{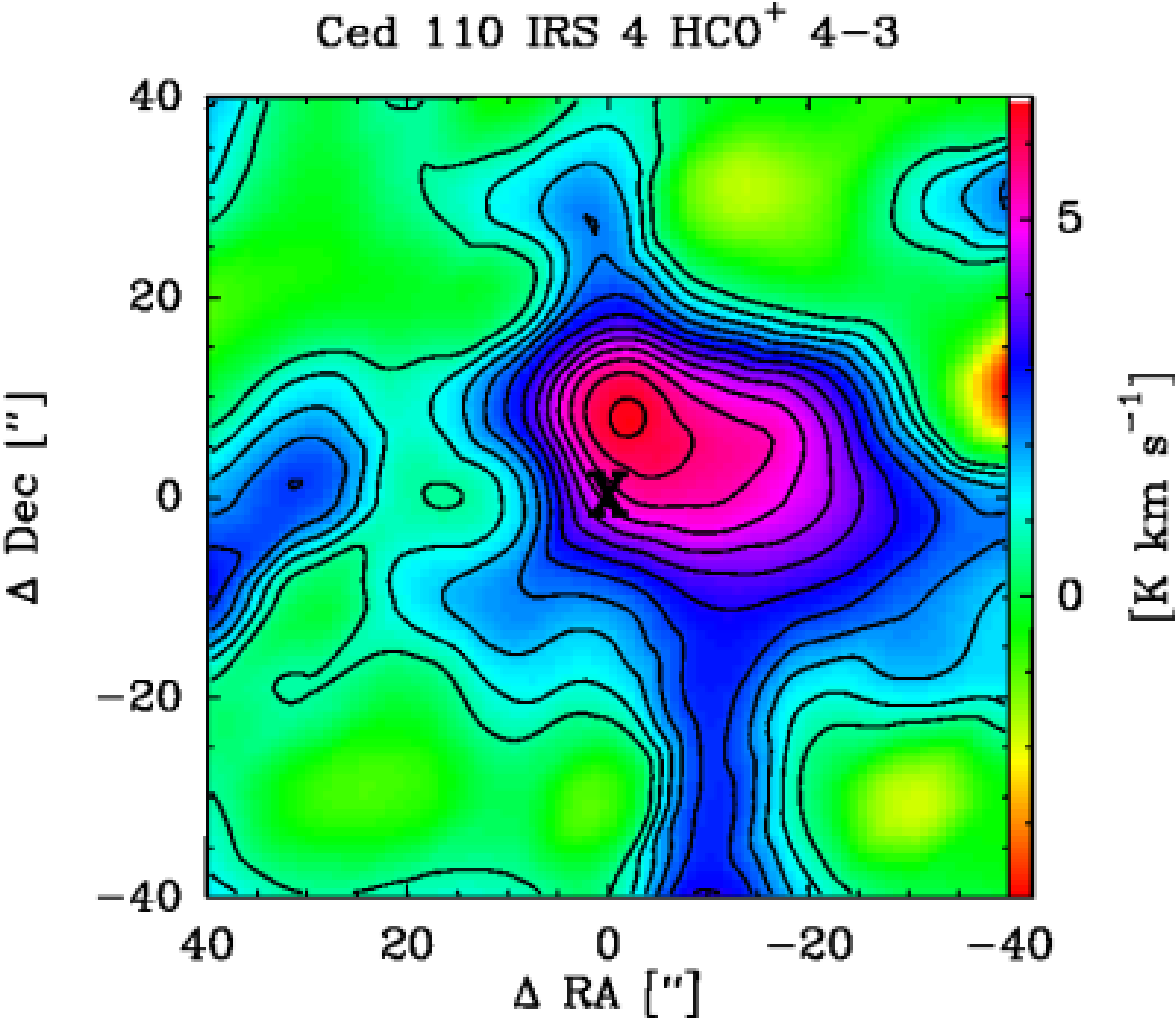}
\includegraphics[width=180pt]{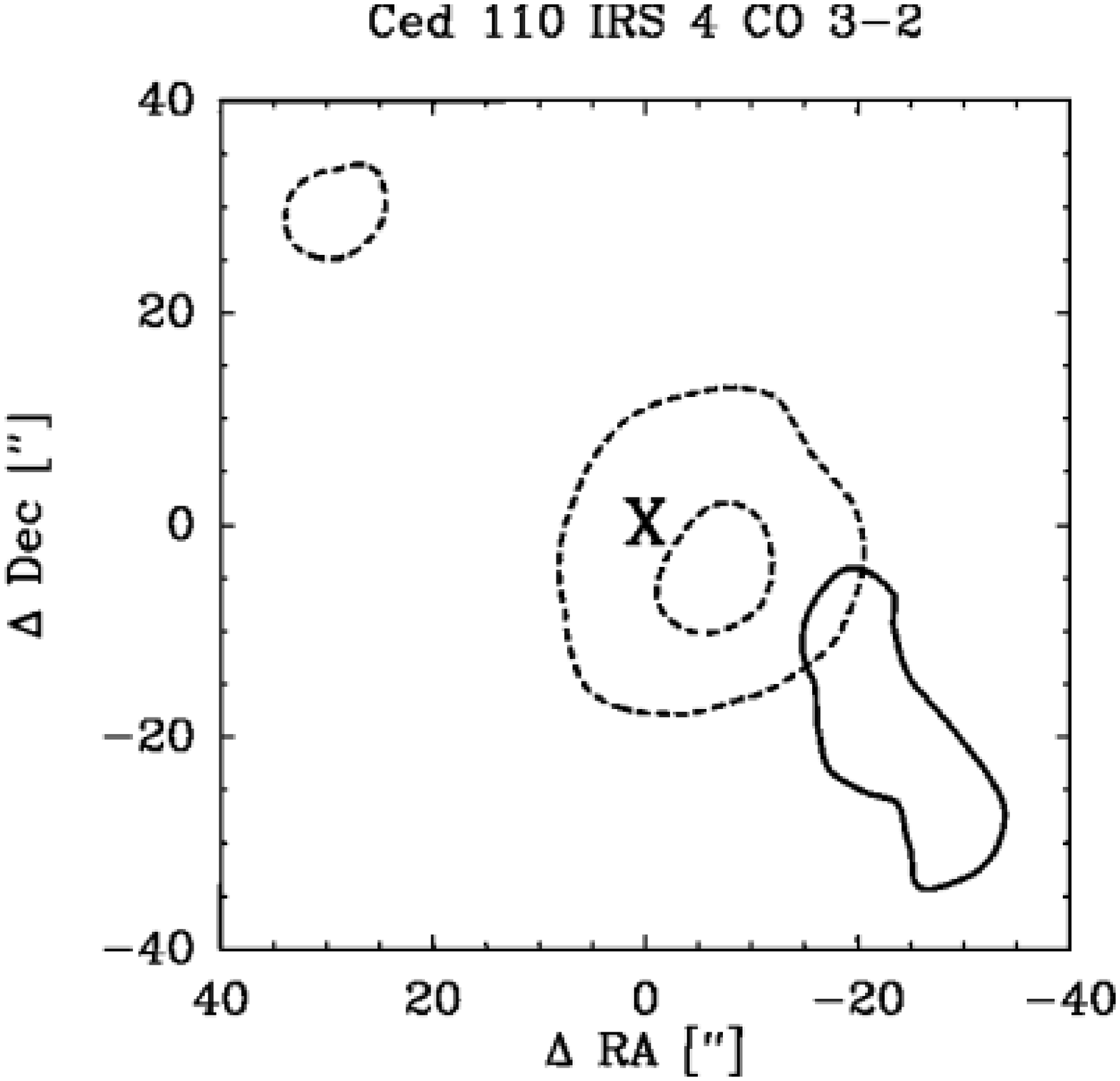}
\includegraphics[width=200pt]{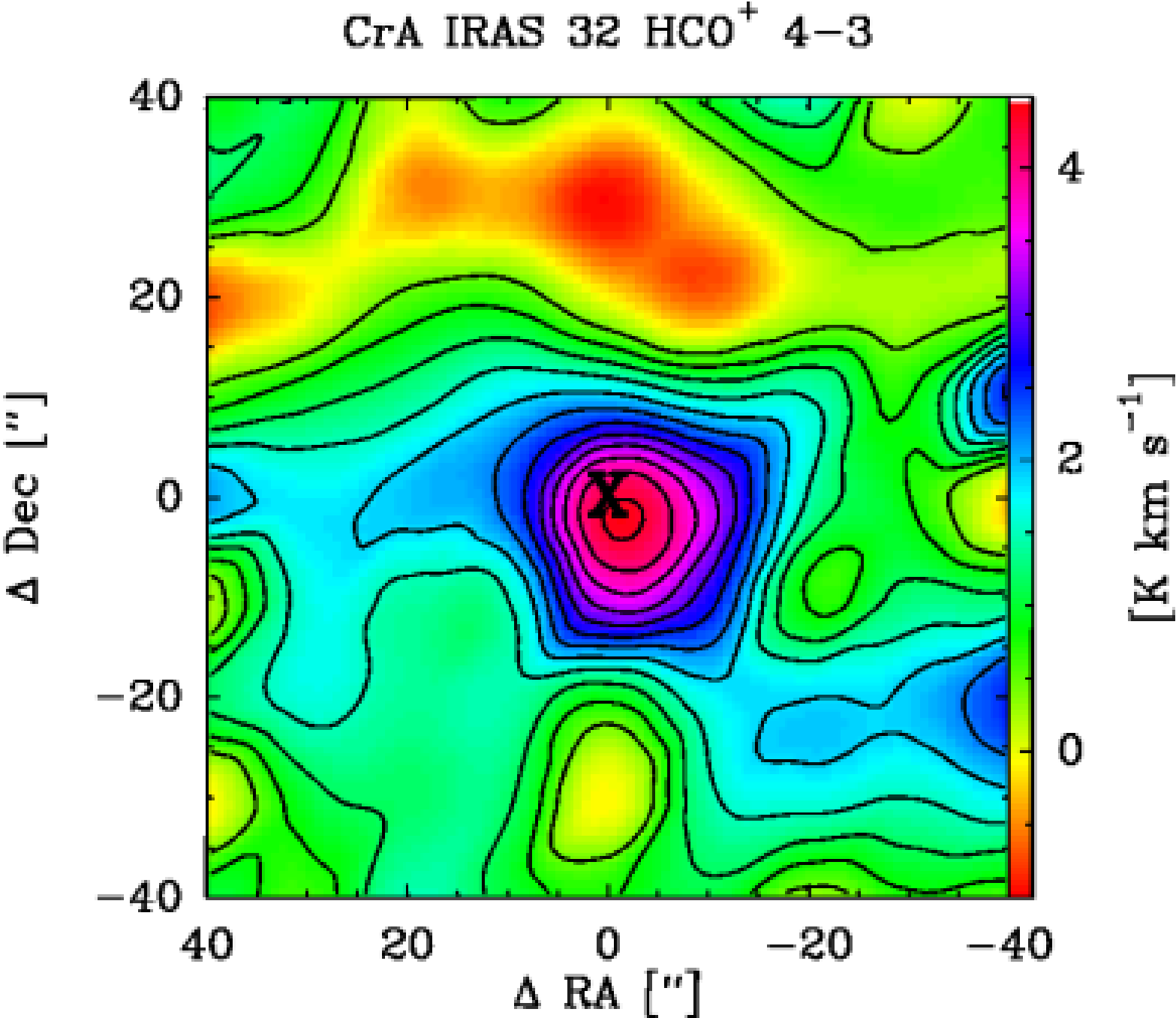}
\includegraphics[width=180pt]{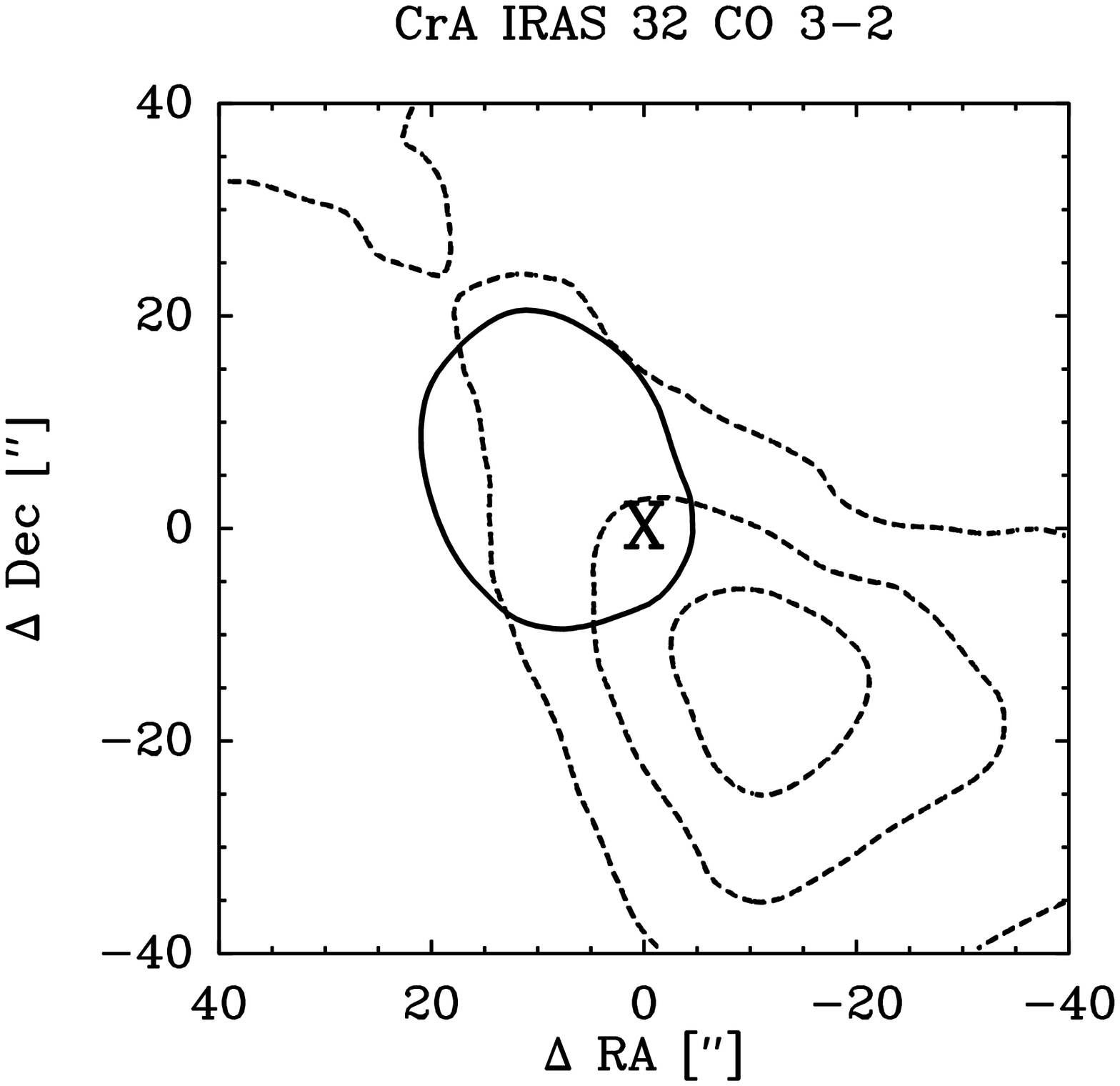}
\includegraphics[width=200pt]{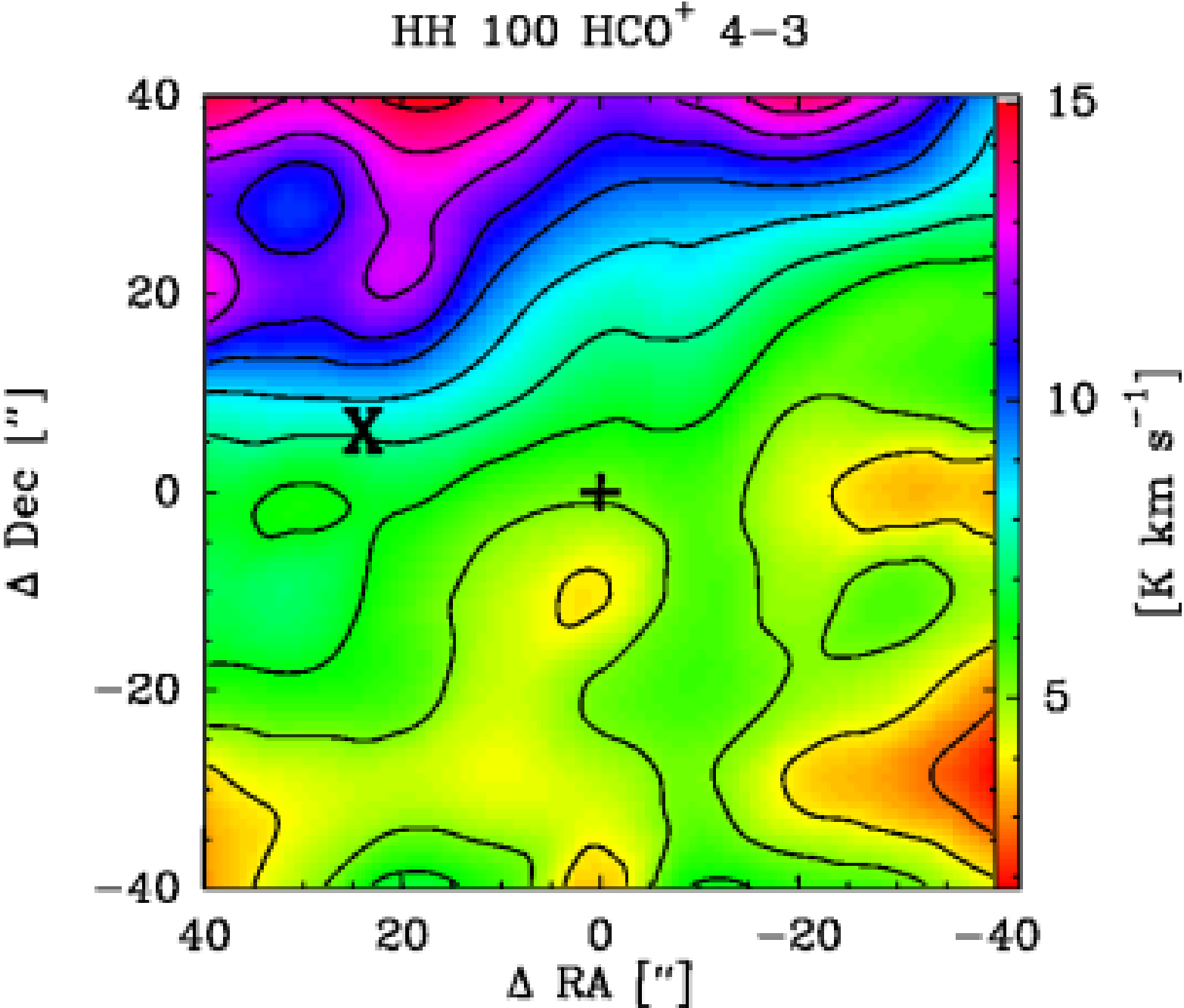}
\includegraphics[width=180pt]{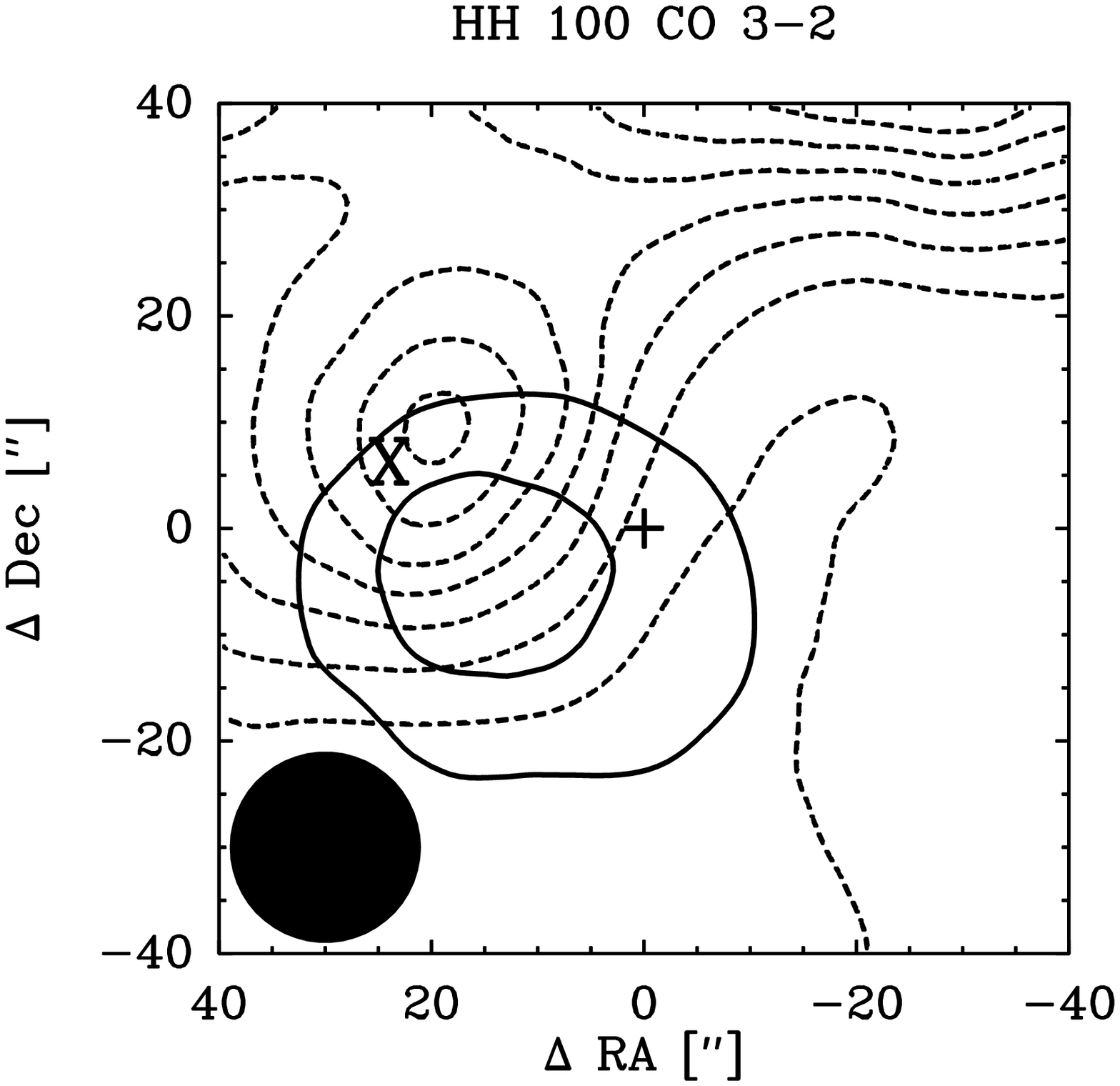}
\end{center}
\caption{HCO$^+$ $J$=4--3 ({\it left}) and CO 3--2 ({\it right}) maps
  of Ced 110 IRS 4 ({\it top}), CrA IRAS 32 ({\it middle}) and HH 100
  ({\it bottom}). The maps are 80$''\times80''$ in size. HCO$^+$ 4--3
  is integrated over the entire line. CO 3--2 shows the red- ({\it
  dashed lines}) and blue-shifted ({\it solid lines}) parts of the
  emission, derived by integrating the line wings of each spectral
  line (velocity difference greater than 1.5 km s$^{-1}$ w.r.t. the
  systemic velocity). Contour levels for the CO 3--2 map are at
  3$\sigma$,6$\sigma$, 9$\sigma$, ... with $\sigma$ equal to the noise
  levels of 0.3 K km s$^{-1}$. The HCO$^+$ contours are
  plotted at 10$\%$, 20$\%$, 30$\%$, ... of the maximum intensity
  given in Table \ref{4:tab:res2ahco}.  The 10$\%$ level corresponds to 7$\sigma$ for Ced 110 IRS4, 4$\sigma$ for CrA IRAS 32 and 9$\sigma$ for HH 100.}  The APEX beam is shown in the
  lower right image. The IR position of HH 100 is marked with a
  $'$X$'$, while the pointed position HH100-off is marked with a plus
  sign. 
\label{4:fig:HH100map}
\end{figure*}
}

\def\placeFigureChapterFourNine{
\begin{figure*}[!ht]
\begin{center}
\includegraphics[width=200pt]{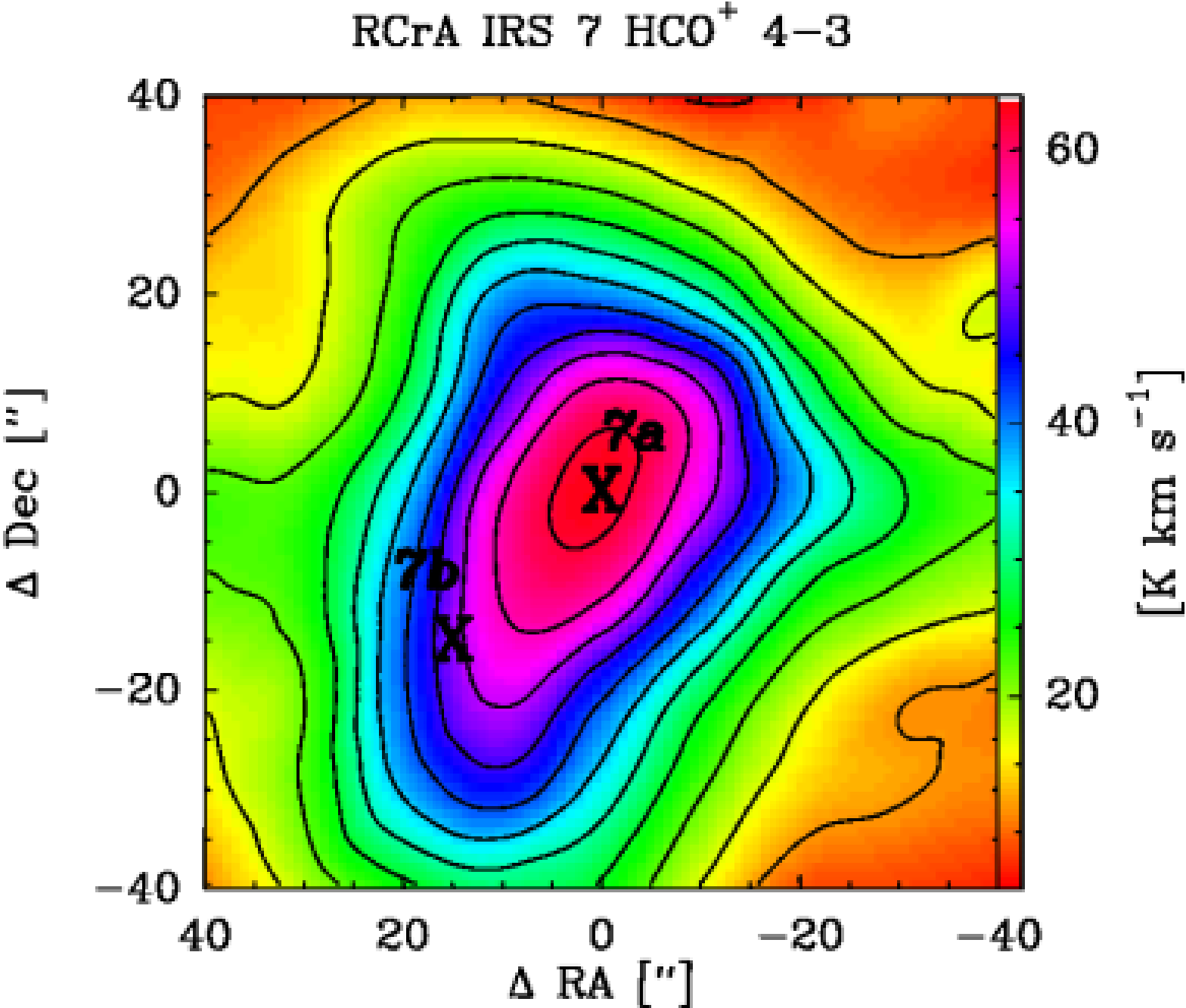}
\includegraphics[width=180pt]{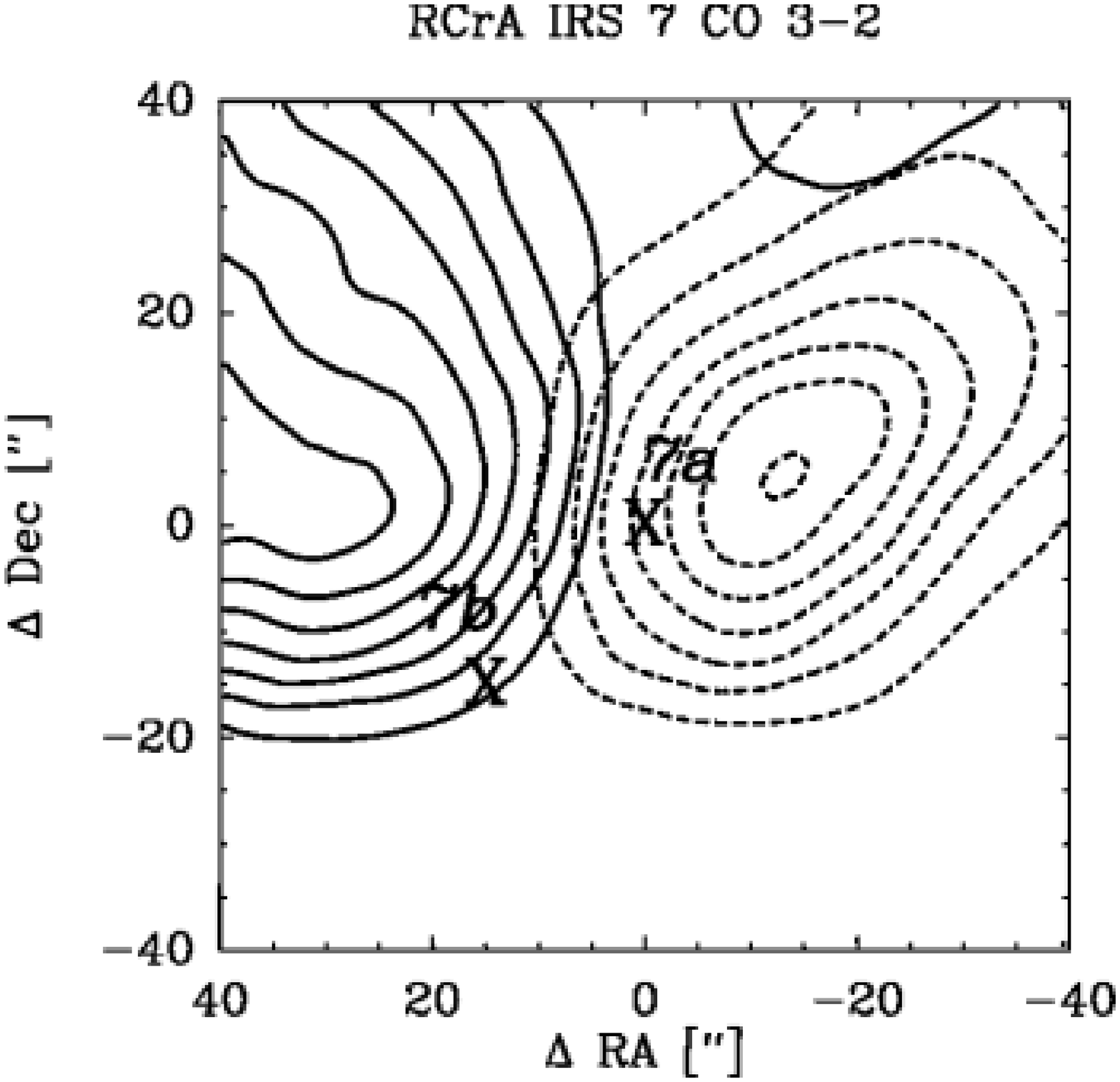}
\includegraphics[width=200pt]{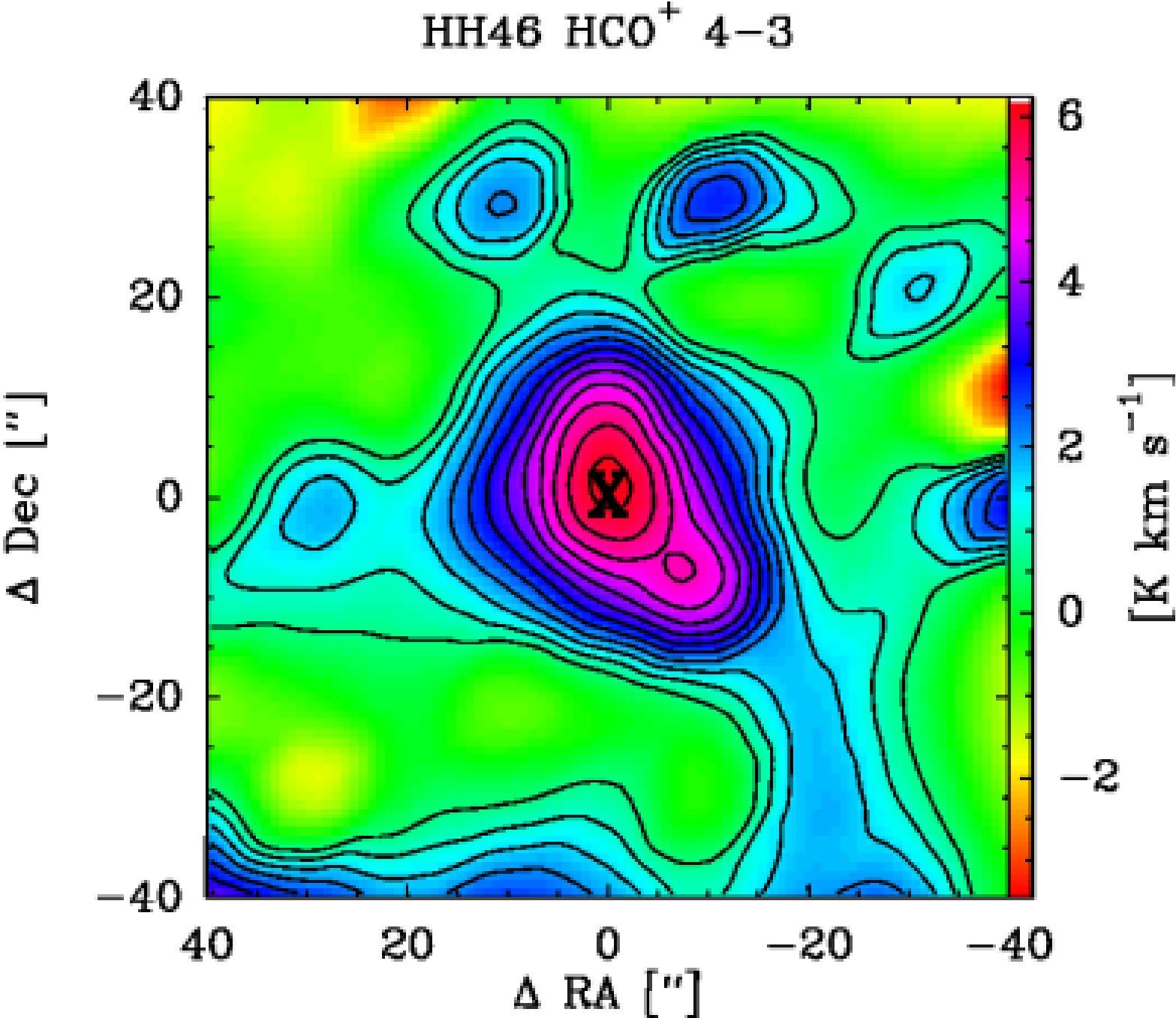}
\includegraphics[width=180pt]{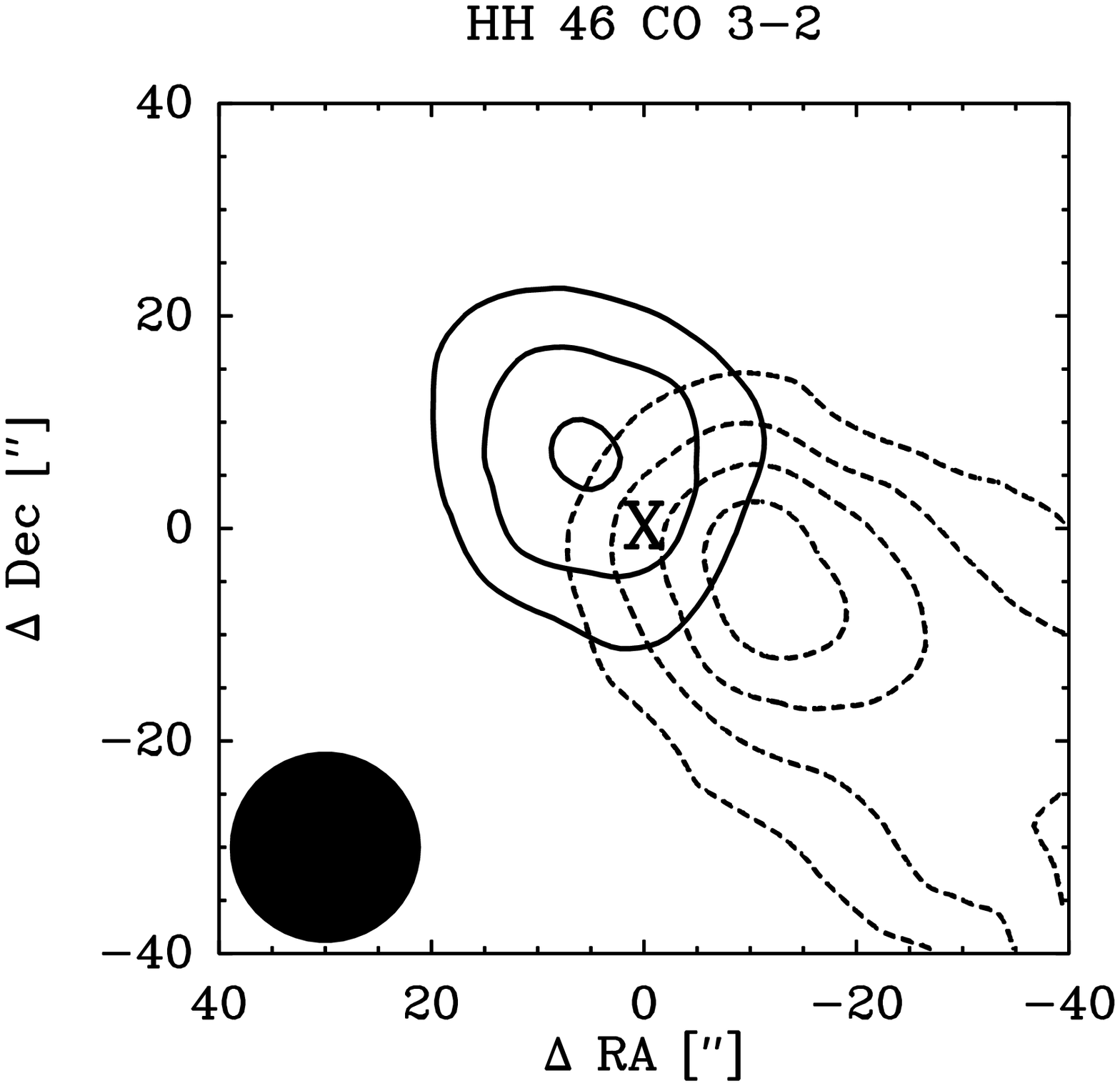}
\end{center}
\caption{HCO$^+$ $J$=4--3 ({\it left}) and CO 3--2 ({\it right}) maps
  of RCrA IRS 7 ({\it top}) and HH 46 ({\it bottom}). The maps are
  80$''\times80''$ in size. HCO$^+$ 4--3 is integrated over the entire
  line. CO 3--2 shows the red- ({\it dashed lines}) and blue-shifted
  ({\it solid lines}) emission, derived by integrating the line wings
  (velocity difference greater than 1.5 km s$^{-1}$ w.r.t. the
  systemic velocity) of each spectral line. Contour levels for the CO
  3--2 map are at 3$\sigma$,6$\sigma$, 9$\sigma$, ... with $\sigma$
  equal to 0.3 K km s$^{-1}$. The HCO$^+$ contours are
  plotted at 10$\%$, 20$\%$, 30$\%$, ... of the maximum intensity
  given in Table \ref{4:tab:res2ahco}.  The 10$\%$ level corresponds to 25$\sigma$ for RCrA IRS 7, and 5$\sigma$ for  HH~46. The APEX beam is shown in the
  lower right image.  Both RCrA IRS 7A and IRS 7B are shown with X in
  the CO map, but the map is centered on RCrA IRS 7A.}
\label{4:fig:HH46map}
\end{figure*}

}

\def\placeFigureChapterFourTen{
\begin{figure*}[!ht]
\begin{center}
\includegraphics[width=130pt]{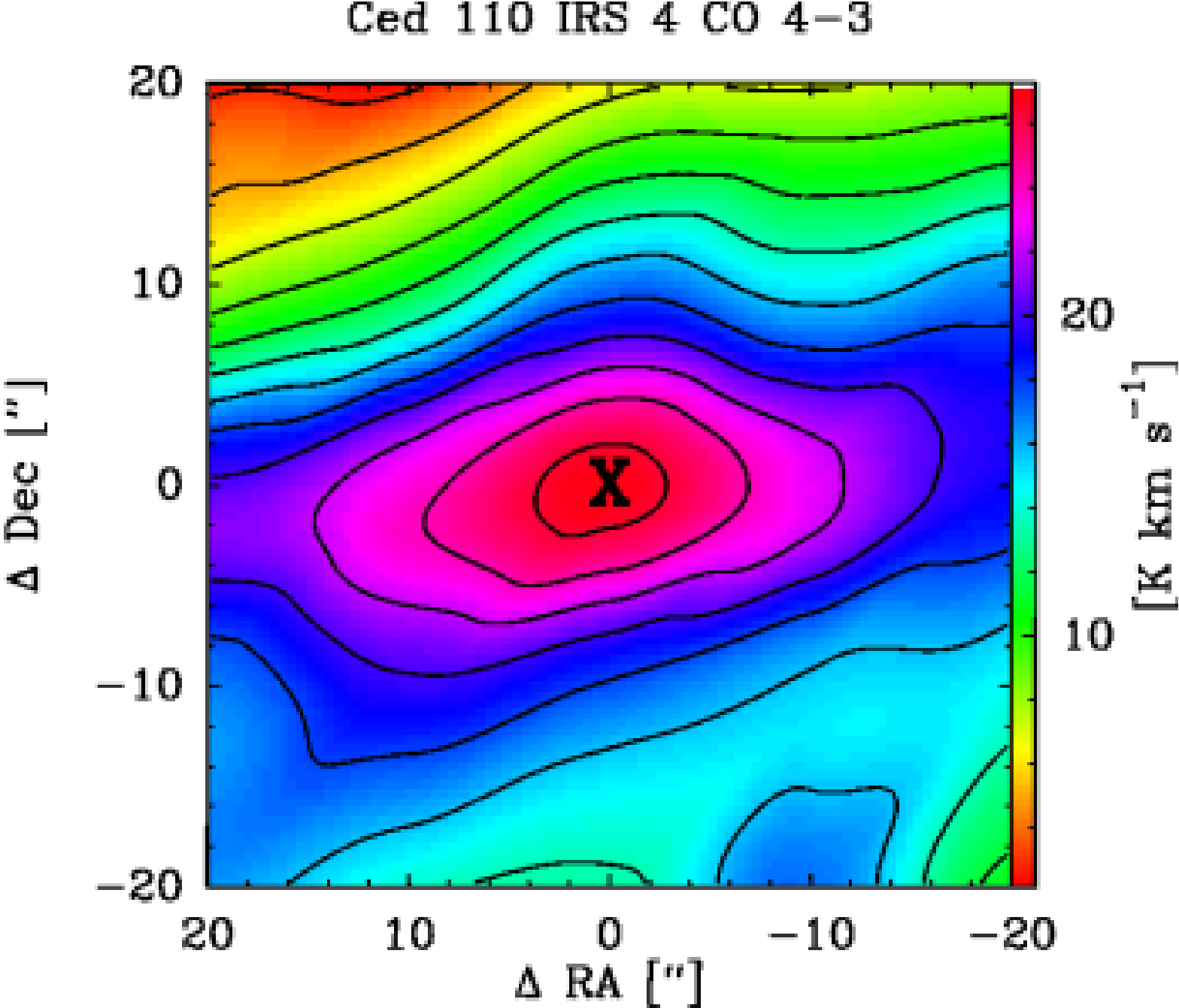}
\includegraphics[width=130pt]{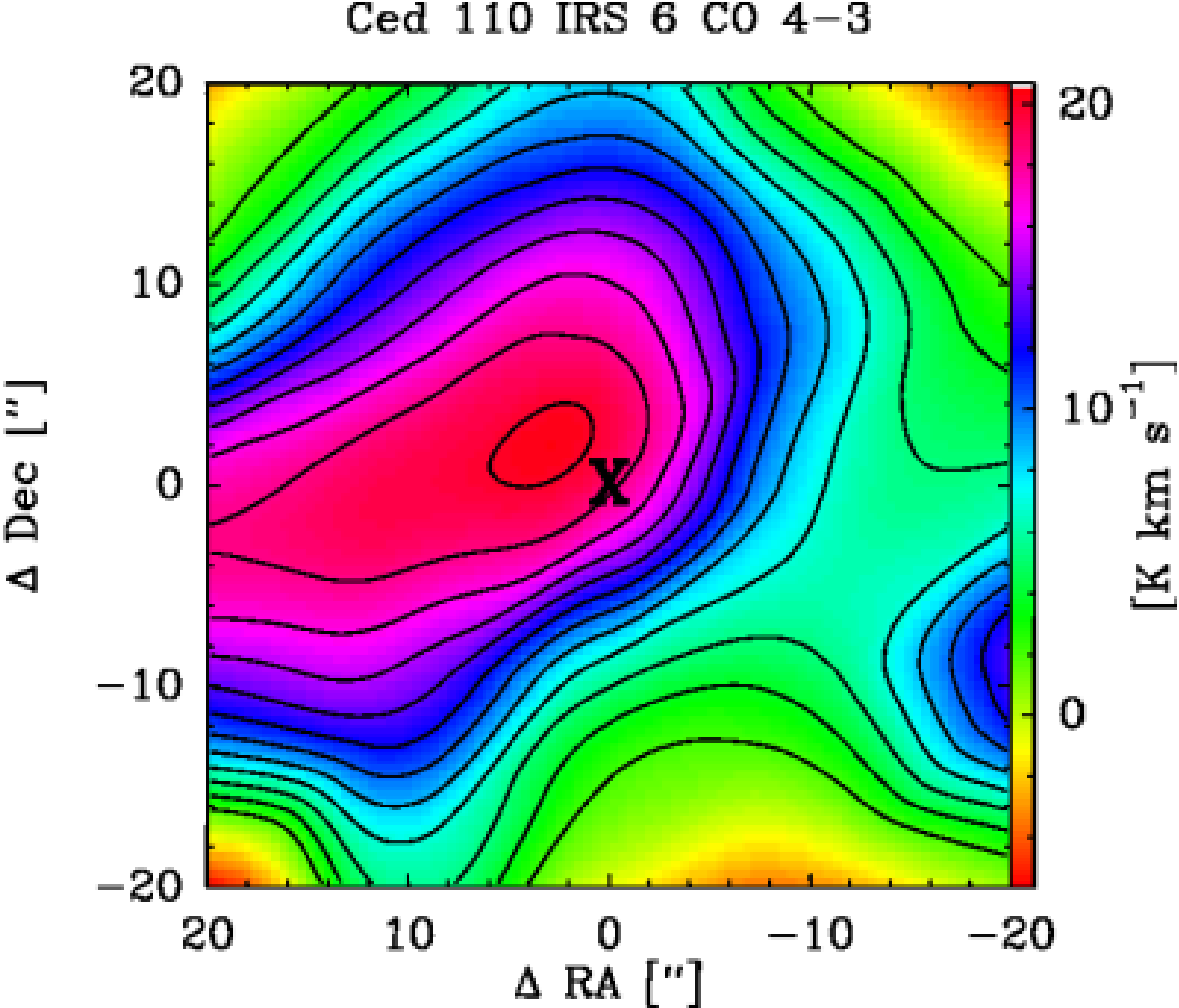}
\includegraphics[width=130pt]{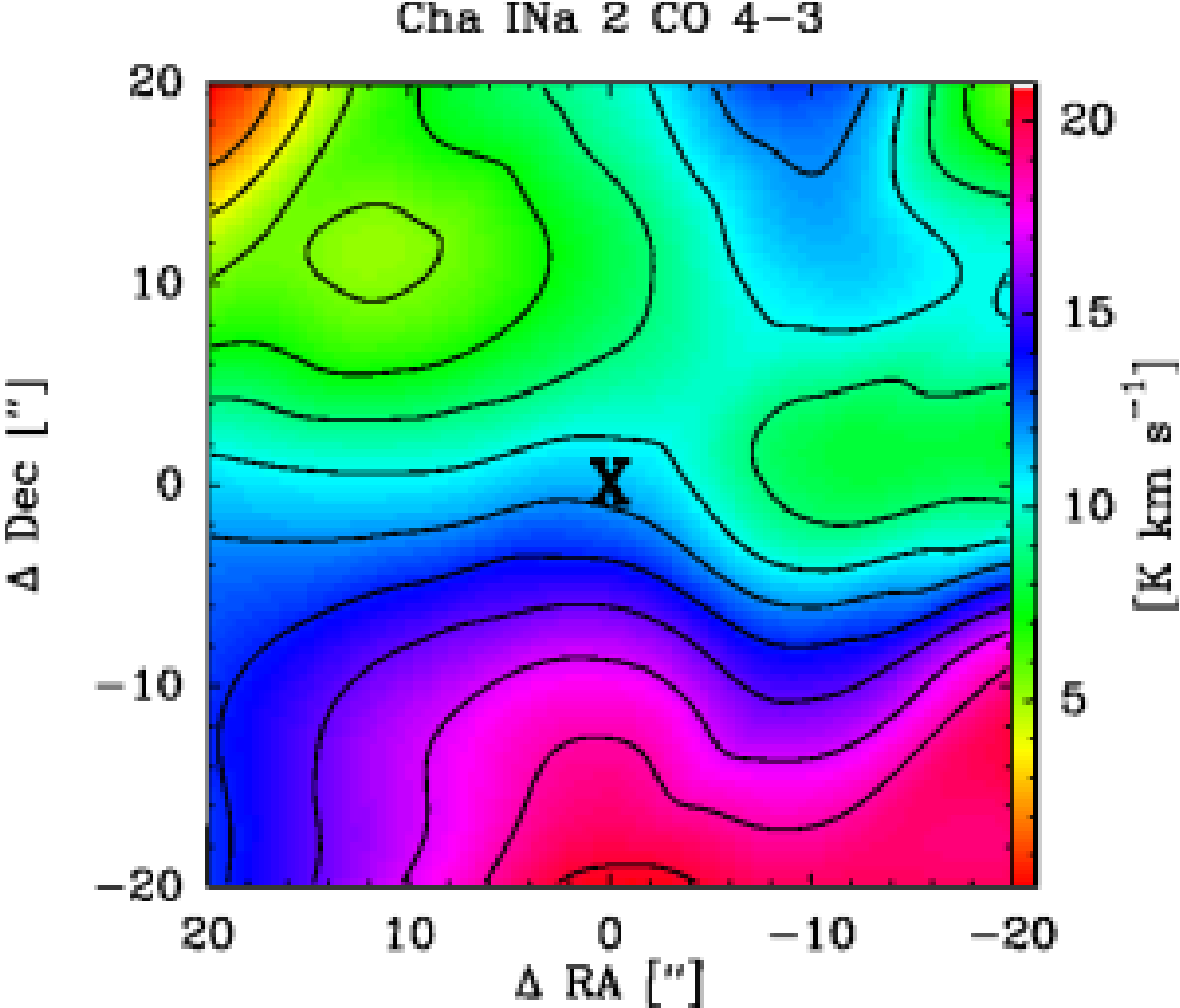}
\includegraphics[width=130pt]{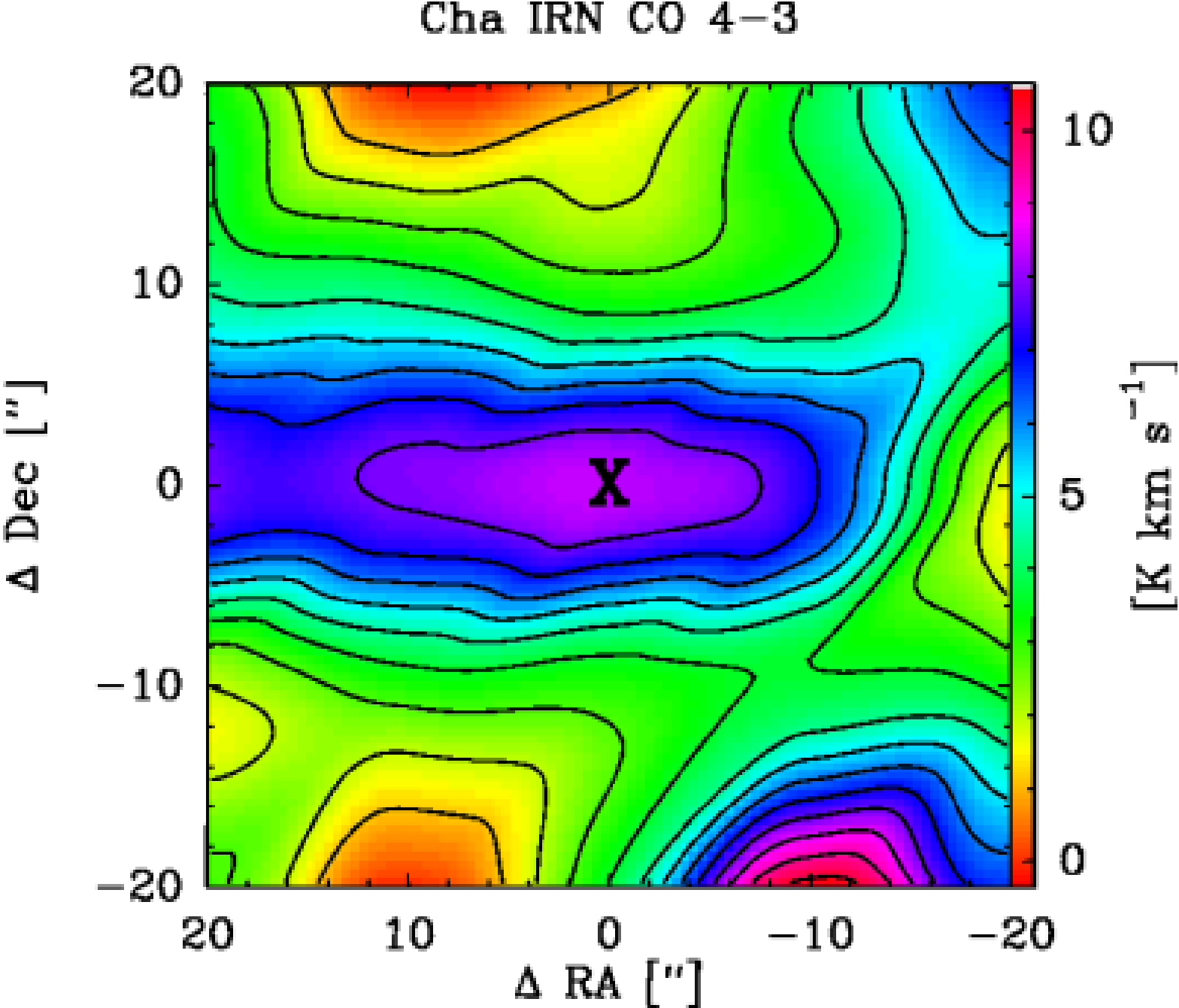}
\includegraphics[width=130pt]{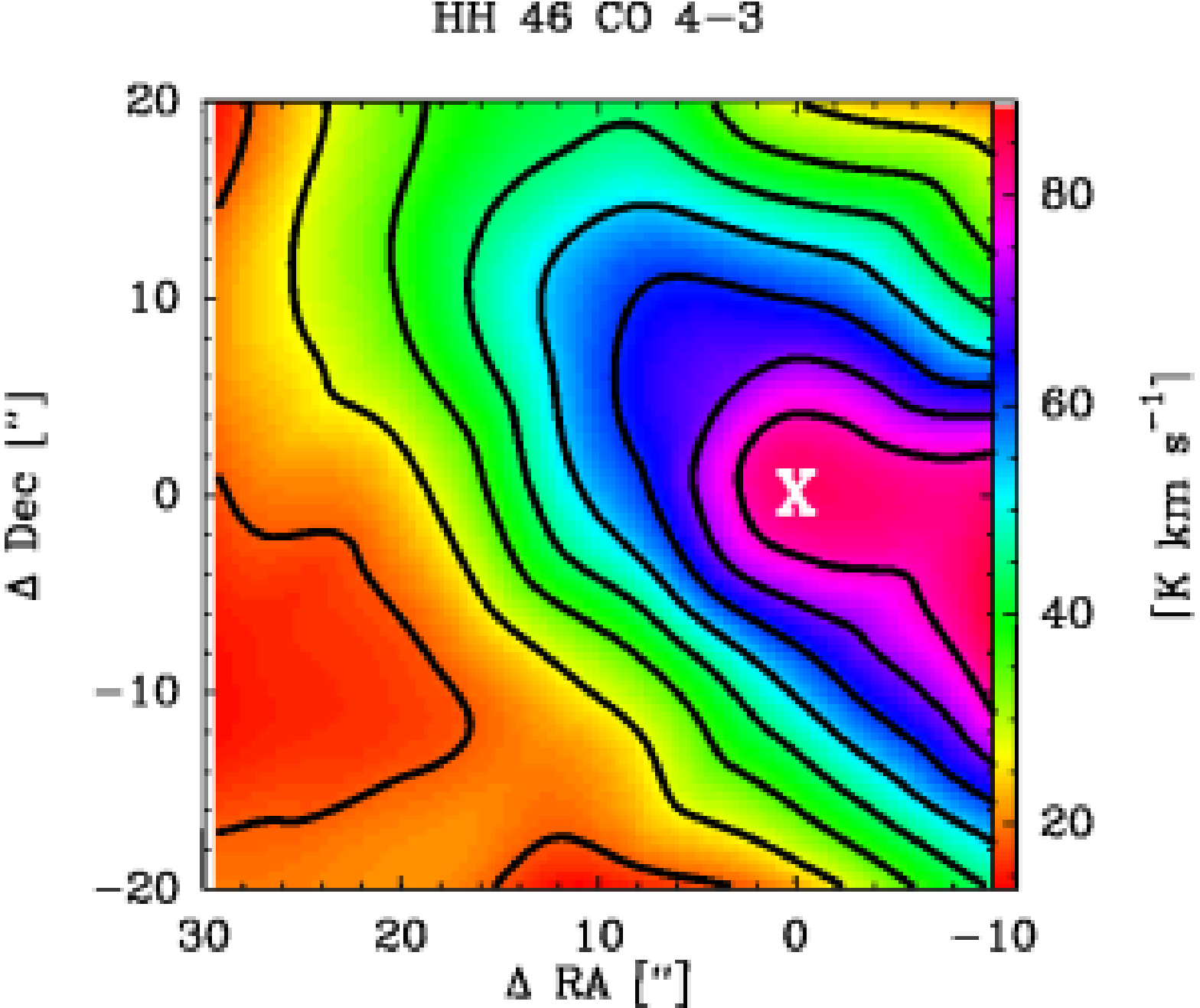}
\includegraphics[width=130pt]{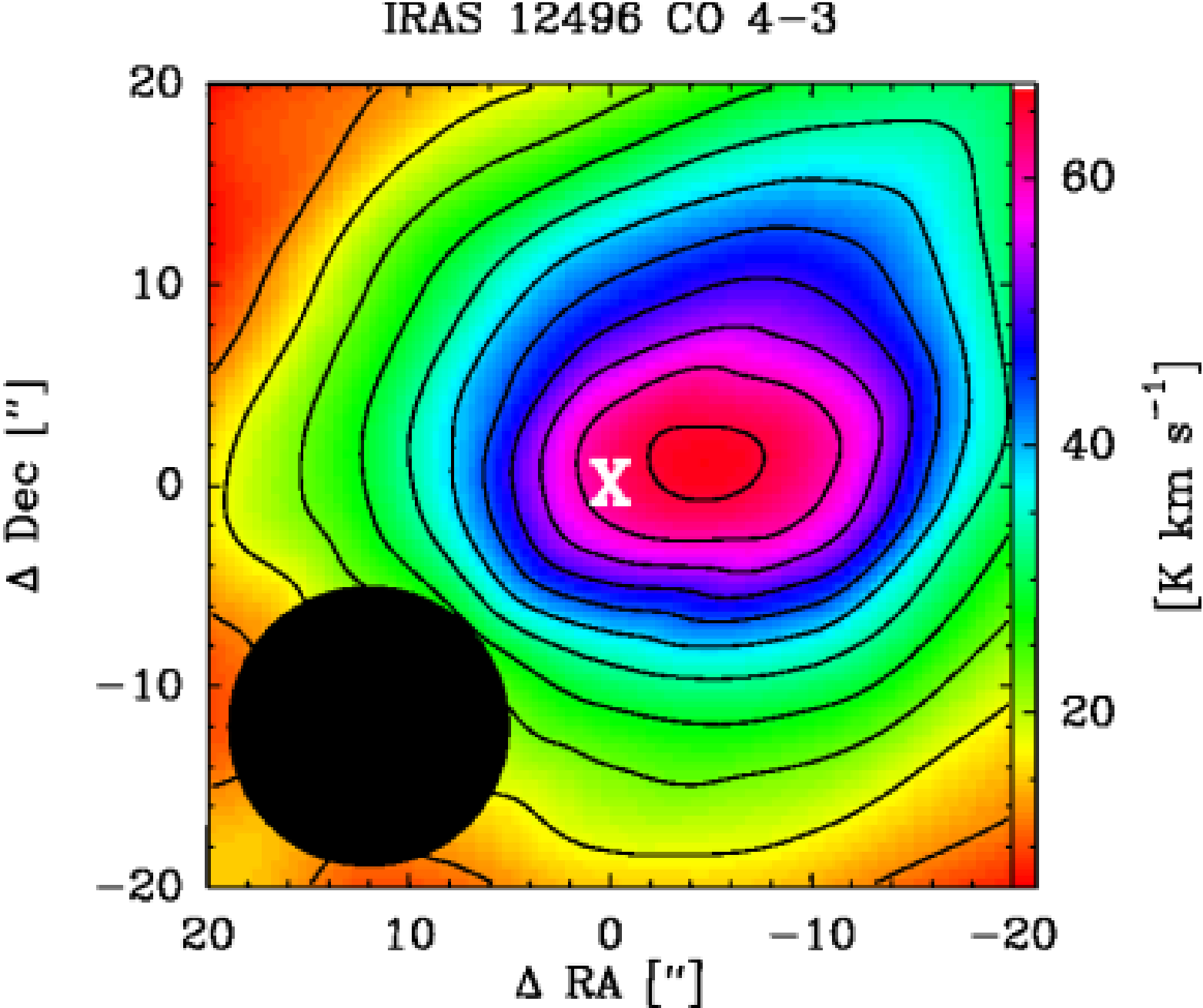}
\end{center}
\caption{CO $J$=4--3 integrated intensity maps taken with FLASH of Ced 110
  IRS 4, Ced 110 IRS 6 and Cha INa2 ({\it top row}), and Cha IRN, HH 46
  and IRAS 12496-7650 ({\it bottom row}). Note that the noise levels
  in these maps are higher than in the spectra given in
  Fig. \ref{4:fig:FLASHCO}. Source positions are marked with 'X'.
    The APEX beam is shown in the lower right image. The lines are drawn to guide the eye.}
\label{4:fig:CO4-3map}
\end{figure*}
}

\def\placeFigureChapterFourIRAS{
\begin{figure*}[!th]
\begin{center}
\includegraphics[width=400pt]{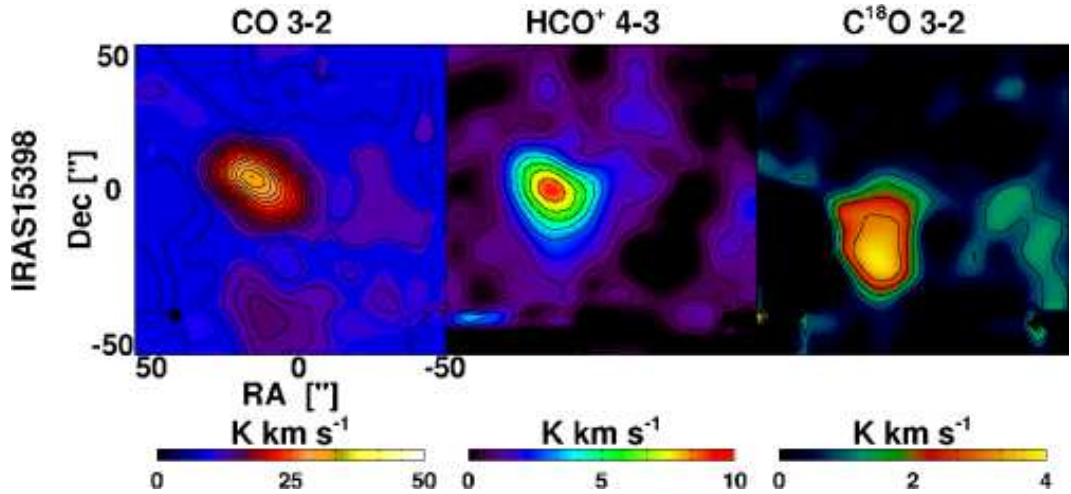}
\end{center}
\caption{The spectrally integrated intensities of CO 3--2 {\it left},
HCO$^+$ 4--3 {\it middle} and C$^{18}$O 3--2 {\it right} observed
toward IRAS 15398--3359 with HARP-B on the JCMT. Spectra were
integrated between velocities of -5 and 15 km s$^{-1}$. The source
itself is at 5.2 km s$^{-1}$. The cores are all slightly resolved,
compared to the 15$''$ beam of the JCMT.}
\label{4:fig:int15398}
\end{figure*}
}

\def\placeFigureChapterFourIRASspec{
\begin{figure}[!th]
\begin{center}
\includegraphics[width=200pt]{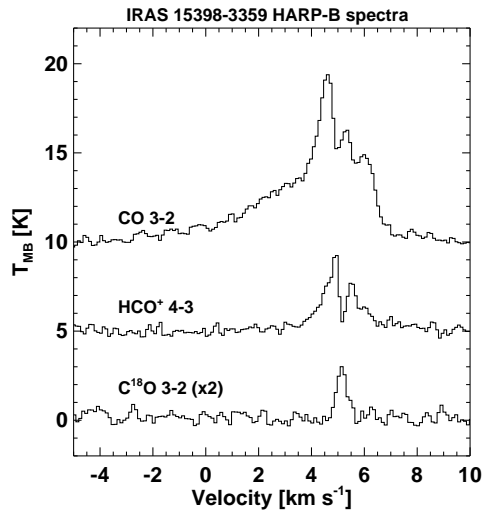}
\end{center}
\caption{The spectra of CO 3--2 {\it top}, HCO$+$ 4--3 {\it middle}
and C$^{18}$O 3--2 {\it bottom} observed with HARP-B on the JCMT at
the central source position.}
\label{4:fig:spec15398}
\end{figure}
}

\def\placeFigureChapterFourEleven{
\begin{figure}[!th]
\begin{center}
\includegraphics[width=240pt]{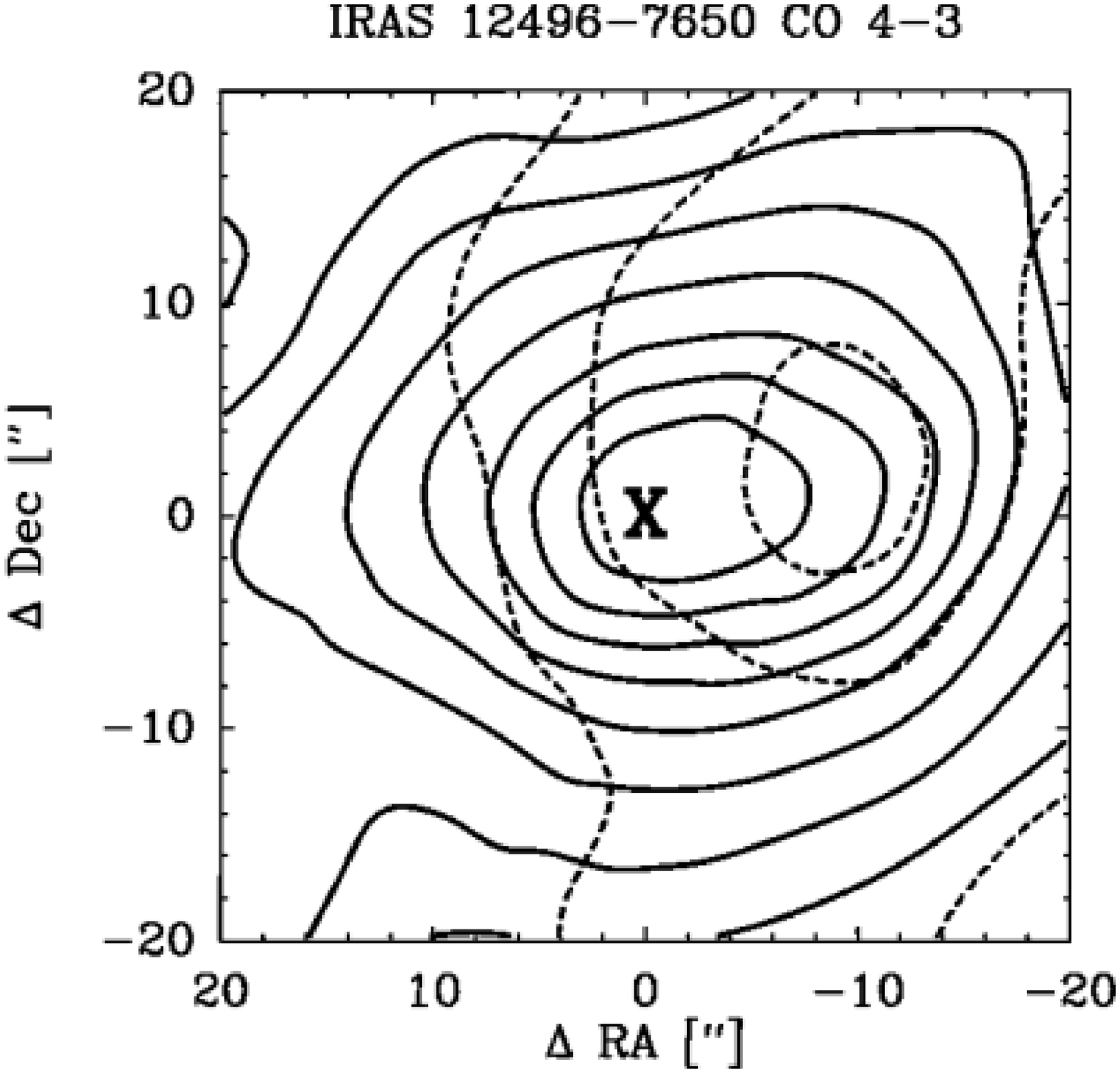}
\end{center}
\caption{CO $J$=4--3 map of IRAS 12496-7650. The blue and red-shifted
outflow emission are shown in {\it solid} and {\it dashed} contours
respectively in steps of 10$\%$ of the maximum of 34.8 K km s$^{-1}$.
}
\label{4:fig:I12496}
\end{figure}
}

\def\placeFigureChapterFourIRASutflow{
\begin{figure}[!th]
\begin{center}
\includegraphics[width=240pt]{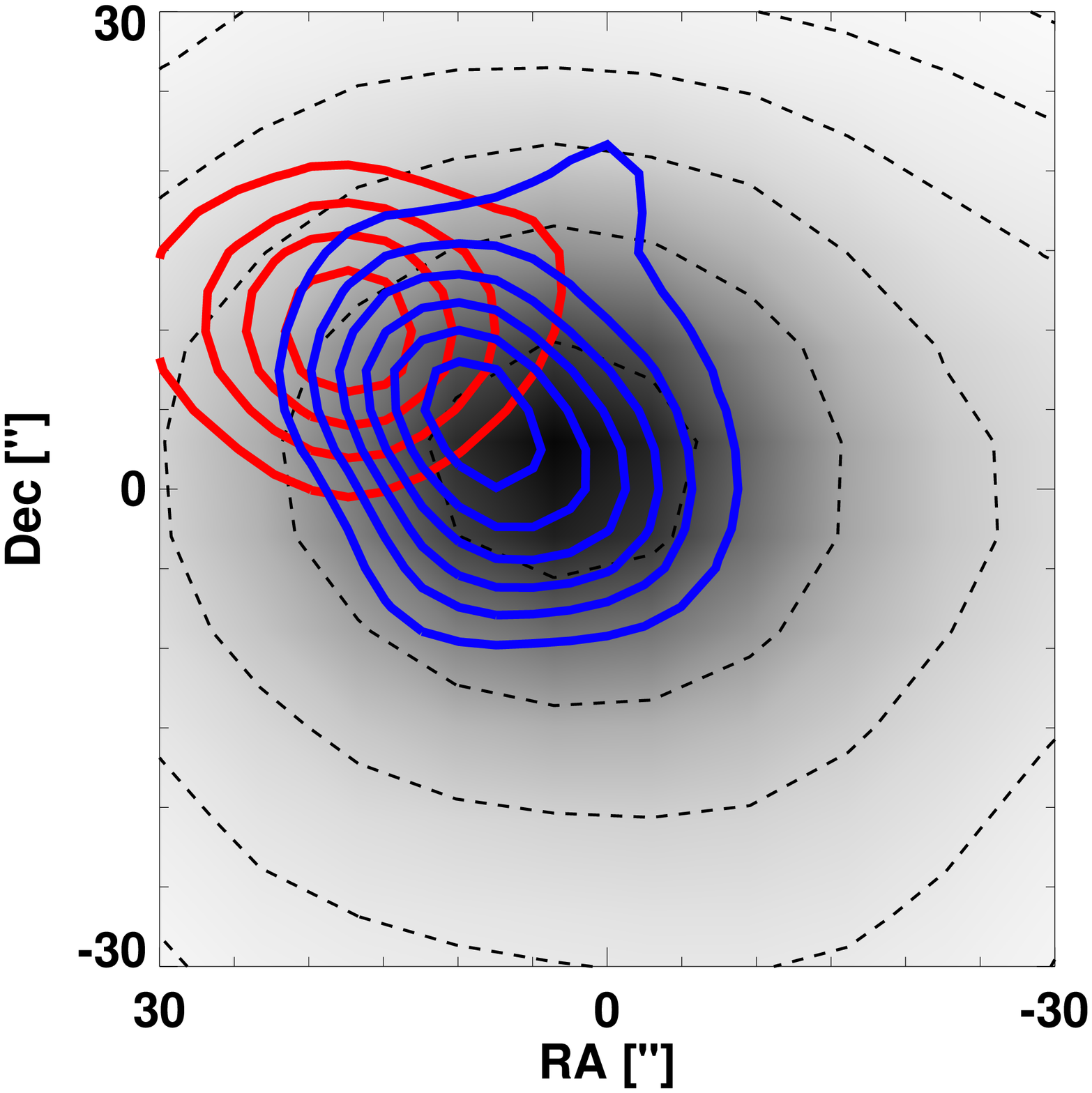}
\end{center}
\caption{CO $J$=3--2 map of IRAS 15398-3359. The blue and red-shifted
outflow emission are shown in {\it blue} and {\it red} contours
respectively in steps of 10$\%$ of the maximum of 8.1 K km s${-1}$. The grey-scale and dashed contour lines in the background are the
850 $\mu$m continuum.}
\label{4:fig:I15398}
\end{figure}
}

\def\placeFigureChapterFourTwelve{
\begin{figure*}[th]
\begin{center}
\includegraphics[width=450pt]{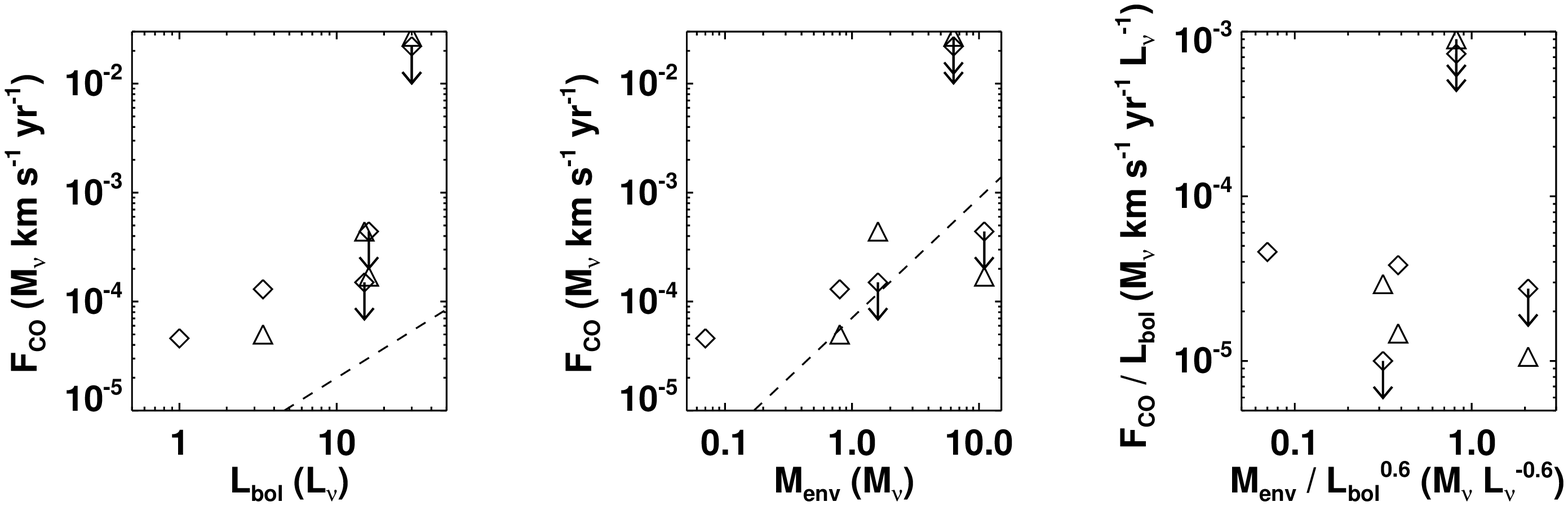}
\end{center}
\caption{CO outflow force $F_{\rm{CO}}$ versus the luminosity,
$L_{\rm{bol}}$ ({\it left}), and versus the envelope mass,
$M_{\rm{env}}$ ({\it middle}). The {\it right} figure shows
$F_{\rm{CO}}$/$L_{\rm{bol}}$ versus
$M_{\rm{env}}$/$L_{\rm{bol}}^{0.6}$, which should be free of most
luminosity and distance effects following Fig. 7 of
\citet{Bontemps96}. Red outflows are shown with diamonds and blue
outflows with triangles. The relations between $F_{\rm{CO}}$,
$L_{\rm{bol}}$ and $M_{\rm{env}}$ found by Bontemps et al.\ are shown
with dashed lines. }
\label{4:fig:forces}
\end{figure*}
}


\def\placeTableChapterFourOne{
\begin{table*}[!t]
\caption{Source sample}
\small
\begin{center}
\begin{tabular}{l l l l l l l}
\hline \hline
Source & RA  & Dec  &  Dist. & Lum. & Ref.$^a$ &Notes$^c$ \\ 
 & (h m s) & (d m s) & (pc) & ($L_{\rm{bol}}$) &  \\ 
 & [J2000] & [J2000] & & & \\\hline
\multicolumn{5}{c}{Chamaeleon I} \\ \hline
Ced 110 IRS 4 & 11 06 47.0 & $-$77 22 32.4 & 150$^e$ &  1   & 1,2,3 & ISO-Cha 84 \\
Ced 110 IRS 6 & 11 07 09.6 & $-$77 23 04.3 & 150 &  0.6 & 1,3 & ISO-Cha 92\\
Cha IRS 6a    & 11 07 09.9 & $-$77 23 06.3 & 150 &  0.8 & 4 &  ISO-Cha 92+ \\
Cha IRN       & 11 08 37.1 & $-$77 43 51.0 & 150 &  5   & 3 &ISO-Cha 150\\
Cha INa 2     & 11 09 36.6 & $-$76 33 39.0 & 150 &  0.6 & 5 &  \\ \hline
\multicolumn{5}{c}{Corona Australis} \\ \hline
RCrA IRS 5  & 19 01 48.0 & $-$36 57 21.6 & 170$^e$& 2   & 6,14 & SMM 5\\
HH 100$^b$  & 19 01 50.7 & $-$36 58 10   & 170& 15  & 6,14 & SMM 3 \\
RCrA IRS 7A & 19 01 55.2 & $-$36 57 21.0 & 170& 3.3$^d$ & 7,14 & SMA 2, SMM 1C\\
RCrA IRS 7B & 19 01 56.2 & $-$36 57 27.0 & 170& 17$^d$  & 7,14 & SMA 1, SMM 1B\\
RCrA TS 3.5 & 19 02 07.1 & $-$36 53 26.2 & 170& -   &  &\\ 
CrA IRAS 32 & 19 02 58.7 & $-$37 07 34.5 & 170& 3.4 & 14 & ISO-CrA 182, SMM 8 \\\hline
\multicolumn{5}{c}{Isolated} \\ \hline
HH 46           & 08 25 43.8 & $-$51 00 35.6 & 450$^f$ &16 &8,9,10 & BHR 36\\
IRAS 07178-4429 & 07 19 21.8 & $-$44 34 55.1 & -   & - & 10,11& CG1/BHR 17 \\
IRAS 12496-7650 & 12 53 17.2 & $-$77 07 10.6 & 200$^g$ &50 & 12& DK Cha, Cham II\\
IRAS 13546-3941 & 13 57 42.2 & $-$39 56 21.0 & 550$^h$ & -  & 10,11 & CG12/BHR 92\\
IRAS 15398-3359 & 15 43 01.3 & $-$34 09 15.0 & 130$^i$ & 0.92& 2,13 & B228, Lupus\\ \hline 
\end{tabular}
\label{4:tab:sou}
\end{center}
$^a$ References: 1. \citet{Luhman08}; 2. \citet{Froebrich05}; 3. \citet{Persi00}; 4. \citet{Persi01};  5. \citet{Persi99}; 6. \citet{Nisini05}; 7. \citet{Groppi07};  8. \citet{NoriegaCrespo04}; 9. \citet{Velusamy07}; 10. \citet{Bourke95}; 11. \citet{Santos98}; 12. \citet{vanKempen06}; 13. \citet{Evans05}; 14. \citet{Nutter05} \\
$^b$ Observations of HH 100 were erronously pointed at coordinates: RA 19h01m49.1s, Dec -36d58m16.0s. See text for further discussion.\\
$^c$ Other names and cloud location.\\
$^d$ The luminosities of the two RCrA IRS 7 sources are not well determined, due to confusion of the total integrated flux at far-IR wavelengths. The total luminosity of both sources (20 $L_\odot$) was used in the characterisation.\\
$^e$ \citet{Knude98} \\
$^f$ \citet{Heathcote96}\\
$^g$ \citet{Hughes92} \\
$^h$ \citet{Maheswar04} \\
$^i$ \citet{Murphy86} \\
\end{table*}
}

\def\placeTableChapterFourTwo{
\begin{table}[!ht]
\caption{Overview of the APEX observations }
\small
\begin{center}
\begin{tabular}{l l l l l}
\hline \hline
Line & &Instrument & Cloud$^a$ & Note \\ \hline
$^{12}$CO & 3--2 & APEX-2a & Cha, CrA, Iso. & \\
 & map & APEX-2a & Various & 80$''\times80''$ \\
& 4--3 & FLASH-I & Cha, Iso & \\
& map & FLASH-I & Cha, Iso & 40$''\times40''$\\
& 7--6 & FLASH-II & Cha, Iso &  \\
& map & FLASH-II & Cha, Iso & 40$''\times40''$ \\
C$^{18}$O & 3--2 & APEX-2a & Cha, CrA, Iso. & \\
HCO$^+$ & 4--3 & APEX-2a & Cha, CrA, Iso. & \\
& map & APEX-2a &Various & 80$''\times80''$ \\
H$^{13}$CO$^+$ &4--3 & APEX-2a & Cha, CrA, Iso. & \\ \hline
\end{tabular} \\
\end{center}
$^a$ Cha = Chamaeleon, CrA = Corona Australis and Iso. = Isolated sample
\label{4:tab:obs}
\end{table}
}

\def\placeTableChapterFourThree{
\begin{table*}[ht]
\caption{HCO$^+$ and H$^{13}$CO$^+$ results. }
\small
\begin{center}
\begin{tabular}{l l l l l l }
\hline \hline
Source & \multicolumn{2}{c}{HCO$^+$ 4--3}  & \multicolumn{2}{c}{H$^{13}$CO$^+$ 4--3} & $V_{\rm{LSR}}$$^a$  \\
   & $\int T_{\rm{MB}} {\rm d}V$$^b$ & $T_{\rm{MB}}$ & $\int T_{\rm{MB}} {\rm d}V$$^b$ & $T_{\rm{MB}}$  \\  
 & (K km s$^{-1}$) & (K) & (K km s$^{-1}$) & (K) & (km s$^{-1}$)\\ \hline
\multicolumn{5}{c}{Chamaeleon} \\ \hline
Ced 110 IRS 4  & 4.0 & 3.1 &  0.38 & 0.25 & 4.3\\
Ced 110 IRS 6  & 0.5 & 0.4 &  $<$0.06 & - & 4.8\\
Cha IRS 6a     & $<$0.2 & -  & $<$0.06 & - & 4.6\\
Cha IRN        & 0.87 & 0.95 &  $<$0.06 & - & 4.5 \\
Cha INa 2      & $<$0.2 &- & $<$0.06 & - & 4.8\\\hline
\multicolumn{5}{c}{Corona Australis} \\ \hline
CrA IRAS 32 &  4.1 & 2.9 &  0.72 & 0.53& 5.6 \\
HH 100-off &  4.8 & 3.2 & $<$0.06 & - & 5.9\\
RCrA IRS 5 &  14.2 & 7.0 &  0.96 & 0.86& 5.7\\
RCrA IRS 7A &  63.7 & 17.4 &  3.8 & 1.6& 5.5\\
RCrA IRS 7B &  49.0 & 18.7 &  3.8 & 2.2& 5.7\\
RCrA TS 3.5 &  $<$0.2 & - &  $<$0.06 & -& 5.9\\\hline
\multicolumn{5}{c}{Isolated} \\ \hline
HH 46 &  8.4 & 5.4 &  1.1& 1.0 & 5.3\\
IRAS 07178-4429 &  1.0 & 0.6 &  $<$0.08 & - & 3.4\\
IRAS 12496-7650 &  1.9 & 1.5 &  $<$0.06 & - & 1.9\\
IRAS 13546-3941 & $<$ 0.12 & -  & $<$0.06 & -& -5.7\\
IRAS 15398-3359$^c$ &  5.9  & 4.6 & - & -& 5.2\\ \hline
\end{tabular}\\
\end{center}
$^a$ The rest velocity $V_{\rm{LSR}}$ was determined by fitting gaussians to either the H$^{13}$CO$^+$ 4--3 (preferred) or C$^{18}$O 3--2 lines (see Table \ref{4:tab:res2aco}).\\
$^b$ Calibration uncertainties estimated at 20$\%$ dominate the uncertainty of the integrated intensity.\\
$^c$ Data from HARP-B map.
\label{4:tab:res2ahco}
\end{table*}
}

\def\placeTableChapterFourFour{
\begin{table*}[ht]
\caption{$^{12}$CO and C$^{18}$O results$^a$}
\small
\begin{center}
\begin{tabular}{l l l l l l l l l l}
\hline \hline
Source & $^{12}$CO 3--2 & & C$^{18}$O 3--2 &  & & CO 4--3 &  & CO 7--6 & \\
   & $\int T_{\rm{MB}} {\rm d}V$$^d$ & $T_{\rm{MB}}$ &  $\int T_{\rm{MB}} {\rm d}V$$^d$ & $T_{\rm{MB}}$ & $\Delta$V &$\int T_{\rm{MB}} {\rm d}V$$^d$ & $T_{\rm{MB}}$& $\int T_{\rm{MB}} {\rm d}V$$^d$ & $T_{\rm{MB}}$\\ 
&  (K km s$^{-1}$) & (K) & (K km s$^{-1}$) & (K)& (km s$^{-1}$)  &  (K km s$^{-1}$) & (K) & (K km s$^{-1}$) & (K) \\ \hline
\hline
\multicolumn{9}{c}{Chamaeleon} \\ \hline
Ced 110 IRS 4   &30.7 & 7.5 & 3.7 & 2.1 & 1.5 & 29.8 & 7.5 & 24.8 & 6.9 \\
Ced 110 IRS 6  & 20.0 & 9.3 & 1.6 & 1.4 & 1.1 & 9.7 & 4.9 & 8.8 & 2.5 \\
Cha IRS 6a  & - & - & 1.2 & 1.1& 1.0 &- & - & - & - \\
Cha IRN & 11.1 & 3.5 & 2.3 & 1.8& 1.2 & 6.5 & 3.0 & 14.0 & 5.2 \\
Cha INa 2 & 11.3 & 6.7 &  1.3 & 1.6 &0.8 & 10.0 & 4.7 & $<$1.5 & - \\\hline
\multicolumn{9}{c}{Corona Australis} \\ \hline
CrA IRAS 32 & 51.6 & 11.0 & 4.6$^e$ & 4.4 & 1.0 &- & - & - & - \\
HH 100-off & 102.6 & 24.7 & 6.3 & 5.2& 1.1 &- & - & - & - \\
RCrA IRS 5 & 144.2 & 29.9 & 11.4 & 8.7& 1.2 &- & - & - & - \\
RCrA IRS 7A & 406.0 & 46.4 & 22.5 & 8.9& 2.2 &- & - & - & - \\
RCrA IRS 7B & 332.7 & 40.4 & 23.0 & 10.0& 2.2 &- & - & - & - \\
RCrA TS 3.5 & 16.8 & 6.4 & $<$0.3 & - & - &- & - & - & - \\\hline
\multicolumn{9}{c}{Isolated} \\ \hline
HH 46$^c$ & 82.5 & 19.4 & 3.2 & 3.3& 0.9 &70.5 & 14.9 & 46.5 & 8.6 \\
IRAS 07178-4429 & 24.0 & 11.4 & 1.7 & 2.3& 0.7  &- & - & - & - \\
IRAS 12496-7650$^b$ &92.8 & 25 & 9.2 & 1.9 & 1.1 &90.0 & 23 & 43.7 & 19.2\\
IRAS 13546-3941 & 16.3 & 11.4 & 1.8 & 2.3& 0.7 &- & - & - & - \\
IRAS 15398-3359 & 25.8 & 9.0 & 0.5 & 1.2 & 0.4  &19.9 &5.6  & 45.3 & 8.5 \\ \hline 
\end{tabular}
\end{center}
\label{4:tab:res2aco}
$^a$ Integrated intensities are across the entire line profile.\\
$^b$ From \citet{vanKempen06}.\\
$^c$ CO 7--6 from \citet{vanKempen09a}.\\
$^d$ Calibration uncertainties estimated at 20$\%$ dominate the uncertainty of the integrated intensity.\\
$^e$ This is a significant discrepancy from the value reported in \citet{Schoeier06}, which determined this intensity to be 6.5 K km s$^{-1}$. A possible explanation can be the inaccurate calibration during science verification of \citet{Schoeier06}.
\end{table*}
}

\def\placeTableChapterFourFive{
\begin{table*}[htp]
\caption{Overview of the envelope properties$^a$.}
\small
\begin{center}
\begin{tabular}{l l l l l l l l}
\hline \hline
Source & $R_{\rm{core}}$ & $M_{\rm{envelope}}$ & $N$(H$_2$)$^a$ & $N$(C$^{18}$O)& $N$(H$^{13}$CO$^+$) & $\tau_{\rm{CO 3-2}}$ & $\tau_{\rm{HCO^+ 4-3}}$  \\ 
 & ($''$) & (M$_\odot$) & (10$^{22}$ cm$^{-2}$) & (10$^{15}$ cm$^{-2}$) & (10$^{15}$ cm$^{-2}$) & & \\ \hline
\multicolumn{7}{c}{Chamaeleon} \\ \hline
Ced 110 IRS 4 & 20 & 0.07 &5.0/-  & 8.3 & 1.4& 181 & 6.7  \\ 
Ced 110 IRS 6 & 20 & 0.04 &3.3/- & 5.5 & - & 90 & $<$13 \\ 
Cha IRS 6a    & - & -    &0.26/- & 0.43 &- & - & - \\ 
Cha IRN       & 20 & 0.17 &4.3/- & 7.2 &- & 400 & $<$5.2\\
Cha INa 2     & - & -    &3.8/- & 6.3 &- & 150 & - \\ \hline
\multicolumn{7}{c}{Corona Australis$^b$} \\ \hline
CrA IRAS 32 &44  & 0.8 &10.5/21.2* & 17.5 & 3.0& 257 & 16 \\
HH100-off   & 45 & 1.6 &12.4/35.2* & 20.7 & - & 130 & $<$0.9\\ 
RCrA IRS 5  & 43 &  1.7 &20.7/24.5* & 35.1 & 4.8 & 189 & 10 \\
RCrA IRS 7A & 48 &  6.3$^c$&21.2/30.5* & 35.3 & 8.9 & 117 & 7.7 \\
RCrA IRS 7B & 48 &  6.3$^c$&23.8/30.5* & 39.7 & 12.3 & 156 & 10\\ 
RCrA TS 3.5 & - & - &-/- & - &  -&  -  & - \\ \hline 
\multicolumn{7}{c}{Isolated} \\ \hline
HH 46$^d$ & 60 & 5.1 & 8.1/18.2* & 13.5 & 6.1 & 103 & 16\\ 
IRAS 07178-4429 & - & - &5.5/- & 9.2 &- & 123.9 & $<$8\\ 
IRAS 12496-7650 & 25 & 0.4 &3.0/- & 5.0 & - & 25 & $<$3.1  \\ 
IRAS 13546-3941 & - & - &5.5/- & 9.2 &- & 123 & - \\ 
IRAS 15398-3359 & 25 & 0.5 &1.9/- & 3.2 & - & 79 & -\\ \hline  
\end{tabular}\\
\end{center}
$^a$ 
The column density of H$_2$ is derived from that of C$^{18}$O assuming
H$_2$/C$^{18}$O=5.5$\times 10^6$. 
If 850 $\mu$m data are available in \citet{diFrancesco08}, column densities 
based on the dust emission are also computed. 
These numbers are indicated as the second
entry with an asterisk. Both column densities are within the same 18 $''$ beam.

$^b$ Mass estimates in the RCrA region may be severely overestimated due to presence of extended cloud material. \\
$^c$ RCrA IRS 7A and 7B cannot be distinguished in the sub-millimeter continuum
maps; the value refers to the core containing both sources. \\
$^d$ For a thorough discussion on the mass of HH 46, see \citet{vanKempen09a}.
\label{4:tab:column}
\end{table*}
}

\def\placeTableChapterFourSix{
\begin{table*}[ht]
\caption{HCO$^+$ 4--3 and CO 4--3 core radii and concentrations}
\small
\begin{center}
\begin{tabular}{l l l l l l}
\hline \hline
Source & $R_{\rm{HCO^+}}$ ($''$) & $C_{\rm{HCO^+}}$ & $R_{\rm{CO(4-3)}}$ ($''$) & $C_{\rm{CO(4-3)}}$ & Emb.$^b$ \\ \hline
\multicolumn{4}{c}{Chamaeleon} \\ \hline
Ced 110 IRS 4 & 21 & 0.77 & 20 &0.69 & y\\
Ced 110 IRS 6 & - & - & 16 &0.74 & y\\
Cha IRS 6a & - & - & - & - & n$^d$ \\
Cha IRN & - & - & 12 & 0.72 & y\\
Cha INa 2 & - & - & - & n.c.$^a$ & n$^d$\\\hline
\multicolumn{4}{c}{Corona Australis} \\ \hline
CrA IRAS 32 & 15 & 0.78  & - & -& y\\
HH 100 & - & n.c.$^a$ & - & -& y$^c$\\
RCrA IRS 5  & - & - & - & - & y$^e$ \\
RCrA IRS 7A & 25& 0.68 & - & - & y \\
RCrA IRS 7B & 25& 0.66 & - & - & y \\ 
RCrA TS 3.5 & - & - & - & - & n$^d$\\ \hline
\multicolumn{4}{c}{Isolated} \\ \hline
HH 46 & 15 & 0.79 &  - & - &y \\
IRAS 07178-4429 & - & n.c.$^a$ & - & - &n \\

IRAS 12496-7650 & - & - & - & - & y$^c$ \\
IRAS 13546-3941 & - & n.c.$^a$  & - & - & n \\ 
IRAS 15398-3359 & 22 & 0.9 & - & - & y \\\hline
\end{tabular}\\
\end{center}
$^a$ n.c. =  no core associated with source \\
$^b$ Our assessment whether the source is embedded or not\\
$^c$ Based on presence of outflow, see  $\S$4.2, 4.3 and Fig \ref{4:fig:HH100map}.\\
$^d$ no HCO$^+$ detected down to the limit.\\
$^e$ Strong HCO$^+$ seen.
\label{4:tab:concen}
\end{table*}
}

\def\placeTableChapterFourSeven{
\begin{table*}[ht]
\caption{Temperature constraints of CO line wings ratios$^d$. }
\small
\begin{center}
\begin{tabular}{l l l l l l l l l}
\hline \hline
Source &   \multicolumn{2}{c}{CO 3--2/4--3} &  \multicolumn{2}{c}{CO 4--3/7--6} & \multicolumn{2}{c}{$T_{\rm{outflow}}$ (K)}& \multicolumn{2}{c}{$T_{\rm{outflow}}$ (K)}\\
 & & & & & \multicolumn{2}{c}{Scenario 1$^a$} &  \multicolumn{2}{c}{Scenario 2$^b$} \\
 & \multicolumn{2}{c}{\hrulefill}  &  \multicolumn{2}{c}{\hrulefill}  &  \multicolumn{2}{c}{\hrulefill}  &  \multicolumn{2}{c}{\hrulefill}\\
 & Red$^c$ & Blue$^c$ & Red & Blue & Red & Blue & Red & Blue  \\ \hline
Ced 110 IRS 4 & 1.0 & 1.2 & 3.5 & -  & 50 & 40 & 170 & 160 \\ 
Ced 110 IRS 6& 2.2 & 1.7 &- & - & $<$30 & 40 & 50 & 60\\ 
HH46 & 1.0 &1.6-2.0 &2.0 & - & 50 & $<$ 40 & 180 & 80 \\ 
IRAS 12496 & - & 0.8 & - & $>$4 & - & 50 & - & $<$200 \\ 
IRAS 15398 & - & 0.7 & - & 0.8 & - & 100 & - & $>$200\\  \hline
\end{tabular}\\
\end{center}
$^a$ Lines thermally excited ($n>n_{\rm{cr}}$) \\
$^b$ Lines subthermally excited ($n$ assumed to be 10$^4$ cm$^{-3}$)\\
$^c$ Red and Blue outflow contributions are calculated by excluding the central 3 km s$^{-1}$ emission.\\
$^d$ Errors on the temperatures are 25$\%$, except for temperatures above 150 K, where it is estimated to be at 40$\%$.\\
\label{4:tab:outflow}
\end{table*}

}

\def\placeTableChapterFourEight{
\begin{table*}[th]
\caption{Outflow parameters from the CO 3--2 mapping. }
\small
\begin{center}
\begin{tabular}{l l l l l l l l l l}
\hline \hline
Source &  $i$ &  Mass$^a$ & $\Delta V_{\rm{max}}$$^b$ & $R_{\rm out}$$^b$ & $t_{\rm{d}}$$^{b,c}$ & $\dot{M}^{a,d}$ & $F_{\rm CO}$$^a$$^{,e}$ & $L_{\rm{kin}}$$^{a,f}$  \\
       &      &(M$_\odot$) & (km s$^{-1}$)         & (AU)    & (yr)              &  (M$_\odot$     &(M$_\odot$ yr$^{-1}$ & (L$_\odot$)  \\ %
       &      &     &           &     &      &  yr$^{-1}$)& km s$^{-1}$) & \\\hline 
\multicolumn{6}{c}{Red Lobe} \\ \hline
Ced 110 IRS 4 & 15  & 1.4e-2& 4.0  & 2.4e3 & 2.8e3 & 4.2e-6 & 4.6e-5 & 7e-5\\ 
CrA IRAS 32   & 45  & 6.6e-2& 6.3 & 9.3e3 & 7.1e3 & 9.3e-6 & 1.3e-4 & 5.5e-4\\
HH 100        & 60  & 3.6e-1& 14.7 & $>$ 6.1e3 & $>$2.0e3 & $<$1.5e-4& $<$4.9e-3 & $<$7.2e-2\\
RCrA IRS 7    & 45  & 1.1e0& 23.0 & $>$1e4& $>$2.1e3 & $<$4.3e-4 & $<$2.2e-2 & $<$5.0e-1 \\
HH46          & 35  & 4.3e-1& 9.5 & $>$2.5e4& $>$1.2e4& $<$3.2e-5 & $<$4.4e-4 & $<$3.1e-3\\  
IRAS 15398-3359& 75 &  2.9e-2& 8.2& 2.9e3 & 4.6e3 & 6.4e-7 & 6.4e-6 & 1.6e-5 \\ \hline
\multicolumn{6}{c}{Blue Lobe} \\ \hline
Ced 110 IRS 4 & 15 & 5.8e-3& -4.2  & - & - & - & - & - \\ 
CrA IRAS 32   & 45 & 1.7e-2& -5.4  & 4.6e3 & 4.0e3& 4.3e-6 & 5e-5 & 1.8e-4\\
HH 100        & 60 & 4.7e-2& -9.7 & 4.6e3 & 2.2e3& 2.1e-5 & 4.4e-4 & 2.9e-4 \\
RCrA IRS 7    & 45 & 1.2e0 & -15.8& $>$5e3& $>$1.5e3& $<$7.9e-4 & $<$2.7e-2 & $<$2.9e-1\\
HH46          & 35 & 1.4e-1& -7.4 & 1.4e4 & 9.0e3& 1.5e-5 & 1.7e-4 & 9.3e-4\\  
IRAS 15398-3359& 75& 9.2e-2& 0.7 & 3.9e3 & 4.1e4& 2.2e-6 & 3.3e-5 & 1.2e-4\\ \hline
\end{tabular}\\
\end{center}
$^a$ Corrected for inclination using the average correction factors of \citet{Cabrit90}. \\
$^b$ Not corrected for inclination. \\
$^c$ Dynamical time scale : $t_{\rm{d}}=R/V_{\rm{max}}$\\
$^d$ Mass outflow rate : $\dot{M}=M/t_{\rm{d}}$.\\
$^e$ Outflow force : $F_{\rm CO}=MV_{\rm{max}}^2/R$ \\
$^f$ Kinetic luminosity : $L_{\rm{kin}}=0.5MV^3_{\rm{max}}/R$\\
\label{4:tab:outflow1}
\end{table*}
}



\thispagestyle{empty}

\abstract { Observations of dense molecular gas lie at the basis of our
  understanding of the density and temperature structure of
  protostellar envelopes and molecular outflows. The 
  Atacama Pathfinder EXperiment (APEX) opens up the study of 
  southern (Dec $<$ -35$^{\circ}$)
  protostars. } {We aim to characterize the properties 
  of the protostellar envelope, molecular
  outflow and surrounding cloud, through observations of high excitation
  molecular lines
  within a sample of 16 southern sources presumed to be embedded YSOs, 
  including
  the most luminous Class I objects in Corona Australis and
  Chamaeleon.}  {Observations of submillimeter lines of
  CO, HCO$^+$ and their isotopologues,
  both single spectra and small maps (up to $80''\times80''$), were
  taken with the FLASH and APEX-2a instruments mounted on APEX to
  trace the gas around the sources. The HARP-B instrument on the JCMT was used to map IRAS 15398-3359 in these lines. HCO$^+$ mapping probes the presence
  of dense centrally condensed gas, a characteristic of
  protostellar envelopes. The rare isotopologues C$^{18}$O and
  H$^{13}$CO$^+$ are also included to determine the optical depth,
  column density, and source velocity. The combination of multiple CO
  transitions, such as 3--2, 4--3 and 7--6, allows to constrain  
  outflow properties, in particular the temperature. Archival
  submillimeter continuum data are used to determine envelope masses.}
{Eleven  of the sixteen sources have associated warm and/or dense ($\geq
  10^6$ cm$^{-3}$) quiescent gas characteristic of protostellar
  envelopes, or an associated outflow.  Using the strength and degree of concentration of
  the HCO$^+$ 4--3 and CO 4--3 lines as a diagnostic,  five sources classified as
  Class I based on their spectral energy distributions are found not
  to be embedded YSOs.
  The C$^{18}$O 3--2 lines show that for none of the sources,
  foreground cloud layers are present.  Strong molecular outflows are found
  around six sources, with outflow forces an order of magnitude
  higher than for previously studied Class I sources of similar
  luminosity. }  {This study provides a starting point for future ALMA
  and Herschel surveys by identifying truly embedded southern YSOs and
  determining their larger scale envelope and outflow
  characteristics.}  \keywords{}
   \maketitle
\section{Introduction}

During the early stages of low-mass star formation, Young Stellar
Objects (YSOs) are embedded in cold dark envelopes of gas and dust,
which absorb the radiation from the central star
\citep{Lada87,Andre93}. This extinction is strong enough that low-mass
embedded YSOs, or protostars, only emit weakly at infrared wavelengths
\citep[e.g.,][]{Jorgensen05,Gutermuth08}. Only at later evolutionary
phases, in which the envelope has been accreted and/or dispersed, does
emission in the optical and infrared (IR) dominate the Spectral Energy
Distribution (SED) \citep[e.g.][]{Hartmann05}. Protostars emit the
bulk of their radiation at far-IR and sub-millimeter wavelengths, both
as continuum radiation, produced by the cold ($T<30$ K) dust, and
through molecular line emission from the gas-phase species present
throughout the protostellar envelope. Although the bulk of the mass
is accreted during the earliest embedded phases, more evolved
protostellar envelopes still contain a reservoir of gas and dust that
can accrete onto the central star and disk system and thus provide the
material for disk and planet formation.  At
the same time, jets and winds from the young star interact with the
envelope and drive molecular outflows which clear the
surroundings. Characterizing and quantifying all of these different
physical components in the protostellar stage is still a major
observational challenge.


The protostellar envelopes and molecular outflows can be directly
  observed either through thermal emission of dust at (sub)millimeter
  wavelengths \citep[e.g.][]{Shirley00,Johnstone01,Nutter05} or the
  line emission of molecules. Low frequency molecular emission traces
  the cold gas in the protostellar envelopes
  \citep[e.g.,][]{Hogerheijde98,Jorgensen02,Maret04} or molecular
  outflows \citep[e.g.,][]{Snell90,Cabrit92,Bachiller99}.
Single-dish observations of dust using current generation bolometer arrays are
able to map large areas and image the surroundings of protostars
\citep[e.g.,][]{Motte98,Shirley00,Stanke06,Nutter08}. 
Through radiative transfer models
\citep[e.g.,][]{Shirley02,Jorgensen02,Young03}, including information
from shorter wavelengths \citep[e.g.,][]{Hatchell07}, the temperature
and density structure of the protostellar envelope can be constrained,
but the continuum data cannot determine the velocity structure of the infalling
envelope, characterize the outflows and their interaction with the
surroundings, or disentangle envelope and (foreground) cloud material.
Analysis of gas observations in the form of spectra of multiple
transitions of the same molecule and its isotopologues provide
additional strong constraints on the physical characteristics of the
protostellar envelope
\citep{Mangum93,Blake94,vanDishoeck95,Schoeier02,Maret04,Jorgensen05,Evans05,vanderTak07}
and outflowing gas
\citep[e.g.,][]{Cabrit92,Bontemps96,Hogerheijde98,Hatchell99,Parise06,Hatchell07a}.

Although many different molecules have been observed in protostellar
envelopes, only a limited number of species are well suited to trace
the physical characteristics of all of the components of an embedded
YSO and its surroundings.  For example, the use of CH$_3$OH and
H$_2$CO is complicated by their changing abundances through the
envelope, although some information can be obtained with careful
analysis and a sufficient number of observed lines
\citep[e.g.][]{Mangum93, Blake94, vanDishoeck95,
  vanderTak00,Leurini04}. The weakness of high-excitation CH$_3$OH and H$_2$CO
lines in all but a handful of the most luminous Class 0 protostars
\citep[e.g.][]{Jorgensen05b} coupled with a lack of some
collisional rate coefficients make these species less suitable tracers for
the bulk of the low-mass embedded YSOs.
In practice, the column density of the surrounding cloud (and that of
any unrelated foreground clouds) is best probed by low excitation
optically thin transitions of molecules with a low dipole moment, such
as the isotopologues of CO, e.g., C$^{18}$O 2--1 or 3--2
\citep{vanKempen09}. For molecular outflows the line wings of various
$^{12}$CO transitions have been efficiently used as tracers of their
properties
\citep{Cabrit92,Bontemps96,Hogerheijde98,Bachiller99,Hatchell99,Hatchell07a}.
The denser regions of the protostellar envelope need to be probed with
emission lines with a high critical density, such as the higher
excitation transitions from HCO$^+$ with critical densities $>10^6$
cm$^{-3}$. The warm gas ($T>$50 K), which can be present in both the
molecular outflow and the inner region of the protostellar envelope,
can only be traced by spectrally resolved high-$J$ CO transitions,
which have $E_{\rm{up}}>50 $ K for $J_{\rm{up}}\geq 4$
\citep{Hogerheijde98}.

Embedded YSOs are generally identified by their 2-24 $\mu$m IR slope
with positive values characteristic of Class I objects \citep{Lada87}.
Over the last several years, it has been found that some of these
Class I objects turn out to be edge-on disks or obscured sources
\citep[e.g.,][]{Luhman99,Brandner00,Pontoppidan05,Lahuis06,vanKempen09}.
The use of a spectral line map over a small ($\sim$2$'\times2'$)
region around the protostar is an elegant solution to disentangle the
contributions of the different components \citep{Boogert02}.  In
particular, the spatial distribution of the HCO$^+$ 4--3 line has
proven to be an excellent diagnostic of such `false' embedded sources:
in the case of Ophiuchus, about 60\% of the sources with Class I or
flat  SED slopes turned out not to be embedded YSOs
\citep{vanKempen09}.

With the development of the Atacama Large Millimeter/Submillimeter
Array (ALMA) on Chajnantor, Chile, surveys of sources in southern
star-forming clouds will be undertaken at high resolution ($\leq$
1$''$) at a wide range of frequencies, and are expected to reveal much
about the inner structure of protostars and their circumstellar
disks. The {\it Herschel Space Observatory} will also target a large
number of low-mass YSOs across the sky.  Both sets of observations
rely heavily on complementary large aperture ground-based single-dish
observations for planning and interpretation.  Apart from providing
the necessary information about the structure on larger scales,
single-dish studies of low-mass star formation in the southern sky
will also put the results obtained from studies on the northern
hemisphere in perspective.

Due to a lack of sub-millimeter telescopes at high dry sites in the
southern hemisphere, high frequency data on southern YSOs are still
limited. The Swedish ESO Submillimeter Telescope (SEST) operated up to
the 345 GHz window, but for only limited periods of the year.  As a
result, southern clouds, such as Chamaeleon, Corona Autralis, Vela or
Lupus, are much less studied than their counterparts in the northern
sky, such as Taurus and Perseus, where such observations have been
readily available for over a decade
\citep[e.g.,][]{Hogerheijde97,Motte98,Hogerheijde98,Johnstone00,Jorgensen02,Nutter05,Nutter08}.
The Atacama Pathfinder EXperiment (APEX) \citep{Guesten06}
\footnote{This publication is based on data acquired with the Atacama
  Pathfinder Experiment (APEX) with programs E-77.C-0217, E-77.C-4010
  and E-78.C-0576. APEX is a collaboration between the
  Max-Planck-Institut f\"ur Radioastronomie, the European Southern
  Observatory, and the Onsala Space Observatory.}  has opened up
access to the atmospheric windows in the 200-1400 $\mu$m wavelength
regime over the entire southern sky.

Of the southern star-forming regions not or only poorly visible with
northern telescopes two regions have proven especially interesting.
The Chamaeleon I cloud ($D$=130 pc, Dec = -77$^\circ$), observed with
{\it IRAS} and the {\it Infrared Space Observatory} (ISO) and included
in the {\it Spitzer Space Telescope} guaranteed time and $`$Cores to
disks' (c2d) Legacy programs 
\citep{Persi00,Evans03,Damjanov07,Luhman08} contains some embedded
sources, especially around the Cederblad region
\citep{Persi00,Lehtinen01,Belloche06,Hiramatsu07}.  Corona Australis
is a nearby star-forming region ($D$=170 pc, see \citet{Knude98}; Dec
= -36$^\circ$), well-known for the central Coronet cluster near the R
CrA star \citep{Loren79,Taylor84,Wilking86,Brown87}. Large-scale
C$^{18}$O 1--0 maps show that it contains about 50 M$_\odot$ of gas
and dust \citep{Harju93}. Many surveys have been undertaken in the IR
\citep[e.g.,][]{Wilking86, Wilking97,Olofsson99,Nisini05}, but only a
few studies mapped this region at submillimeter wavelengths
\citep{Chini03,Nutter05}. Although Corona Australis has only a few
protostars with rising infrared SEDs, the cloud does contains some of
the most luminous low-mass protostars in the neighborhood of the Sun
($D<$ 200 pc) with luminosities of up to 20 L$_{\rm{\odot}}$.

In this paper we present observations of submillimeter lines of CO and
HCO$^+$ of a sample of 16 embedded sources in the southern sky,
with a focus on the Chamaeleon I and Corona Australis clouds, to
identify basic parameters such as column density, presence and
influence of outflowing material, presence of warm and dense gas and
the influence of the immediate surroundings, in preparation for
Herschel and ALMA surveys or in-depth high resolution interferometric
observations.  In section $\S$ 2 we present the observations, for
which the results are given in $\S$ 3 with a distinction between the
clouds and isolated sources. Both the single spectra ($\S$ 3.1) and
maps ($\S$ 3.2) are presented.  In $\S$ 4 we perform the analysis of
the observations, making use of archival submillimeter continuum data,
with the final conclusions given in $\S$ 5.

\placeTableChapterFourOne

\section{Technical information}
\subsection{Source list}

Sources were selected from a sample of sources with rising infrared
SEDs (Class I) observed with the InfraRed Spectrograph (IRS) on {\it
  Spitzer} in the scope of the c2d legacy program \citep{Evans03} and
with the Very Large Telescope (VLT) of the European Southern
Observatory \citep{Pontoppidan03} (Table~1). At the time of selection
of the {\it Spitzer} and VLT sources in 2000, the list included the
bulk of the known southern low-mass Class I YSOs with mid-infrared
fluxes of at least 100 mJy.  Most of these sources are located in the
Chamaeleon and Corona Australis clouds, supplemented by a few other
isolated Class I sources found initially by IRAS.
Within the Chamaeleon I cloud,
most sources are located in the Cederblad region
\citep{Persi00,Hiramatsu07}. Cha IRS 6a is located 6$''$ away from Ced
110 IRS 6 ($\sim$half an APEX beam at 460 GHz).
 Ced 110 IRS 4 was studied by \citet{Belloche06} which detect an outflow with an axis of $<$30$^\circ$, which makes this source a prime target in Chamaeleon. 
 The embedded YSOs in the Corona
Australis region are mostly located in the small region around R CrA,
called the Coronet \citep{Wilking86,Wilking97,Nutter05}, but two
sources are located in the R CrA B region.  RCrA IRS 7A and B are believed to be two embedded objects in a wide binary configuration with a separation of 30$''$ \citep{Nutter05,Schoeier06}. These are deeply embedded as little to no IR counterparts were found \citep{Groppi07}. HH~100 is in a region that is dominated by the RCrA A complex, but is an embedded source in itself according to \citet{Nutter05}. CrA IRAS 32 is located in R CrA B region, but has little to none known surrounding material.
HH 46 is a Class I protostar famous for its well-characterized outflow, pointed to the south-west
\citep{Chernin91,Heathcote96,Stanke99,NoriegaCrespo04,Velusamy07}. It is extensively discussed in \citet{vanKempen09a}. IRAS
12496-7650, also known as DK Cha,  is one of the brightest protostars in the southern
hemisphere and a potential driving source of HH 54
\citep{vanKempen06}. ISO-LWS observations were detected with ISO \citep{Giannini99}, but not confirmed by \citet{vanKempen06}. IRAS 07178-4429, IRAS 13546-3941 and IRAS
15398-3359, all very interesting embedded sources in the southern hemisphere, are also included
\citep[e.g.,][]{Bourke95,Shirley00,Haikala06}.  Table \ref{4:tab:sou}
lists the sources of our sample and their properties.  Luminosities
are adopted from the c2d lists \citep{Evans03,Evans08} and the VLT
lists \citep{Pontoppidan03}.

 During the reduction, it was discovered that the
 original position of HH 100 was incorrect and differed by
 $\sim$20$''$ from the IR position. Throughout this paper, the
 incorrect position will be referred to as 'HH 100-off', while 'HH
 100' will refer to the actual source. Single-pixel spectra for HH 100-off
are shown in some of the figures, as they are useful as probes of the cloud
 material and outflows in the RCrA region, whereas the true HH 100 source 
 position
 is included in the maps.

\placeTableChapterFourTwo
\placeFigureChapterFourOneTwo
\placeFigureChapterFourThreeFour

\subsection{Observations}

Observations of the sources in  Table \ref{4:tab:sou}
were carried using the APEX-2a \citep[345
GHz,][]{Risacher06} and the First Light APEX Submillimeter Heterodyne
\citep[460/800 GHz, denoted as FLASH hereafter,][]{Heyminck06})
instruments mounted on APEX between August 2005 and July 2006. FLASH
allows for the simultaneous observation of molecular lines in the 460
GHz and 800 GHz atmospheric windows.

\placeFigureChapterFourFive
\placeFigureChapterFourSix

\placeFigureChapterFourSeven

Using APEX-2a, observations were taken of the $^{12}$CO 3--2 (345.796
GHz) , C$^{18}$O 3--2 (329.331 GHz), HCO$^+$ 4--3 (356.734 GHz) and
H$^{13}$CO$^+$ 4--3 (346.998 GHz) molecular lines of the entire sample
at the source position with the exception of Cha IRS 6a in
  $^{12}$CO. The $^{12}$CO 4--3 (461.041 GHz) and 7--6 (806.652 GHz)
transitions were observed with FLASH for the Chamaeleon I sample as well as some of the isolated sample (Ced
110 IRS 4, Ced 110 IRS 6, Cha INa 2, Cha IRN, HH46, IRAS 12496-7650
and IRAS 15398-3359).  The single spectra of CO 3--2, C$^{18}$O 3--2,
CO 4--3 and 7--6 for IRAS 12496-7650 can be found in
\citet{vanKempen06}, but the integrated intensities are included here
for completeness. Similarly, HH 46 is discussed more extensively in
\citet{vanKempen09a}. Typical beam sizes of APEX are
18$''$, 14$''$ and 8$''$ at 345, 460 and 805 GHz with respective main
beam efficiencies of 0.73, 0.6 and 0.43 \citep{Guesten06}.

For both instruments, new Fast Fourier Transform Spectrometer (FFTS)
units were used as back-ends, with over 16,000 channels available
\citep{Klein06}, allowing flexible observations up to a resolution of
60 kHz (0.05 km s$^{-1}$ at 345 GHz). Pointing was checked using line
pointing on nearby point sources and found to be within 3$''$ for the
APEX-2a instrument. The pointing of FLASH was not as accurate with
excursions up to 6$''$ for these observations. However, it is not expected 
that this has significant influences on the results, as most spectra were 
extracted from the peak intensities of the maps. Calibration was done on hot and
cold loads as well as on the sky. Total calibration errors are of the
order of 20$\%$. Reference positions of 1,800$''$ to 5,000$''$ in
azimuth were selected. For RCrA IRS 7 and 5, emission
in the CO 3--2 line at this position was subsequently found resulting
in artificial absorption in the spectra, but this did not exceed 5
$\%$ of the source emission and is less than the calibration
uncertainty. 

Observations at APEX were taken under normal to good weather conditions with
the precipitable water vapor ranging from 0.4 to 1.0 mm.  Typical
system temperatures and integration times were 600 and 1000 K and 5 
and 4 minutes for APEX-2a and FLASH-I, respectively. All spectra at
a given position were summed and a linear baseline was subtracted
within the CLASS package.  The APEX-2a observations
were smoothed to a spectral resolution of 0.2 km s$^{-1}$. The RMS was
found to be 0.15-0.2 K in 0.2 km s$^{-1}$ channels for HCO$^+$ 4--3,
$^{12}$CO 3--2 and C$^{18}$O 3--2. H$^{13}$CO$^+$ spectra were
re-binned to 0.5 km s$^{-1}$ for deriving the (upper limits on)
integrated intensity. The RMS for the CO 4--3 observations is 0.35 K
and for CO 7--6 1.4 K in channels of 0.3 km s$^{-1}$, significantly
higher than the APEX-2a observations because of the higher system
temperatures.

In addition to single spectra, small (up to 80$''\times80''$) fully
sampled maps were taken using the APEX-2a instrument (see Table
\ref{4:tab:obs}) of $^{12}$CO 3--2
and HCO$^+$ 4--3 for HH 46, Ced 110 IRS 4, CrA IRAS
32, RCrA IRS 7 and HH 100 . With FLASH, similar maps
were taken of the Cha I sample, HH 46 and IRAS 12496-7650 in $^{12}$CO
4--3 and 7--6, but not of the RCrA sample. All maps were taken in on
the fly (OTF) mode and subsequently rebinned to 6$''$ . The RMS in the maps is a factor of 5 or more higher
than that in the single spectra. For example, for the CO 4--3 maps
taken with FLASH the noise levels are $\sim$0.9 K in a 0.7 km s$^{-1}$
channel, compared with 0.35 K in a 0.3 km s$^{-1}$ for the single
position spectra.

 For IRAS 15398-3359, maps were also taken with the 16-pixel
 heterodyne array receiver HARP-B mounted on the James Clerk Maxwell
 Telescope (JCMT)\footnote{The James Clerk Maxwell Telescope is
 operated by The Joint Astronomy Centre on behalf of the Science and
 Technology Facilities Council of the United Kingdom, the Netherlands
 Organisation for Scientific Research, and the National Research
 Council of Canada. Data were obtained with programs M06BN11, M07BN09
 and M08AN05} in the CO 3--2, C$^{18}$O 3--2 and HCO$^{+}$ 4--3
 lines. The high spectral resolution mode of 0.05 km s$^{-1}$
 available with the ACSIS back-end was used to 
 disentangle foreground material from outflowing material.  Spectra
 were subsequently binned to 0.15 km s$^{-1}$. The HARP-B pixels have
 typical single side-band system temperatures of 300--350 K. The 16
 receivers are arranged in a 4$\times$4 pattern, separated by
 30$''$. This gives a total footprint of 2$'$ with a spatial
 resolution of 15$''$, the beam of the JCMT at 345 GHz.  The
 2$'\times2'$ fields were mapped using the specifically designed
 jiggle mode HARP4\footnote{See JCMT website
 http://www.jcmt.jach.hawaii.edu/}.  Calibration uncertainty is
 estimated at about 20$\%$ and is dominated by absolute flux
 uncertainties. Pointing was checked every two hours and was generally
 found to be within 2--3$''$. The map was re-sampled with a pixel size
 of 5$''$, which is significantly larger than the pointing error.  The
 main-beam efficiency was taken to be 0.67. The JCMT data for IRAS
 15398-3359 are presented in $\S$ 3.3 \\

All APEX data were reduced using the CLASS reduction package
\footnote{CLASS is part of the GILDAS reduction package. See
http://www.iram.fr/IRAMFR/GILDAS for more information.}. For the JCMT
data the STARLINK package GAIA and the CLASS package were used. \\

\placeFigureChapterFourEight
\placeFigureChapterFourNine

\placeFigureChapterFourTen

\placeTableChapterFourFour

\placeTableChapterFourThree

\section{Results}
\subsection{APEX Single spectra at source position}
Spectra are shown in Figs. \ref{4:fig:COC18O} - \ref{4:fig:isoHCO}
(all spectra taken with APEX-2a), and Fig. \ref{4:fig:FLASHCO}
(spectra taken with FLASH). The integrated intensities and peak
temperatures at the source positions are given in Tables
\ref{4:tab:res2aco} (CO and C$^{18}$O observations with APEX-2a and
FLASH observations) and \ref{4:tab:res2ahco} (HCO$^+$ and
H$^{13}$CO$^+$ with APEX-2a).  Source velocities are determined with
an accuracy of 0.1 km s$^{-1}$ by fitting gaussian profiles to the
optically thin lines of H$^{13}$CO$^+$ (preferred if detected) or
C$^{18}$O. The results are presented in Table
\ref{4:tab:res2ahco}. Table \ref{4:tab:res2aco} also includes the
width of the C$^{18}$O 3--2 line, if detected. These are generally
narrow, of order 1 km s$^{-1}$.

In general the 3--2 and 4--3 lines of $^{12}$CO cannot be fitted with
single gaussians, because of self-absorption at or around the source
velocity $V_{\rm{LSR}}$. This absorption is especially deep in RCrA
IRS 7A and 7B with $\tau >> 1$. Due to this absorption, the total
integrated line strength of $^{12}$CO spectra is not useful in the
analysis of the quiescent molecular gas. Only for a few sources,
e.g. IRAS 07178-4429, can the $^{12}$CO 3--2 be fitted with a single
gaussian.  Interestingly, the self-absorption is absent in all three
detected CO 7--6 lines in Ced 110 IRS 4, Cha INa 2 and Cha IRN at the
$S/N$ level of the observations.  All C$^{18}$O 3--2 lines can be
yfitted with single gaussians, indicating that there are no unrelated
(foreground) clouds along the line of sights. The C$^{18}$O 3--2 lines
are much stronger in the R CrA cloud than in the other clouds.

The majority of the sources show strong line wings, a sign of a
bipolar outflow. Especially the sources in RCrA have prominent
emission up to 15 km s$^{-1}$ from the line center.

The HCO$^+$ lines, which can be used to trace the dense gas, 
cover the widest range of antenna temperatures and include a variety
of line profiles. As two extreme cases, no HCO$^+$ 4--3 is seen toward
Cha INa2 down to 0.2 K, whereas the spectrum for RCrA IRS 7A peaks at
$T_{\rm{MB}}$=18 K. Note that the brightest spectra in Corona
Australis all show self-absorption, indicating that the HCO$^+$ 4--3
is optically thick at the positions of the envelope. 
Ced 110 IRS 4, RCrA IRS 7A and 7B and CrA IRAS 32 have line profiles
suggesting dense outflowing gas.  The H$^{13}$CO$^+$ 4--3 line is only
detected for 5 sources,
although the detection of Ced 110 IRS 4 is only 3.4$\sigma$.

\placeTableChapterFourFive

\subsection{APEX maps}
The maps of the total integrated intensities of HCO$^+$ 4--3 can be
found in Fig. \ref{4:fig:HH100map} and \ref{4:fig:HH46map}, alongside
outflow maps of $^{12}$CO 3--2.  In the outflow maps, the integrated
spectrum is split between red-shifted ({\it dashed contour}),
quiescent ({\it not shown}) and blue-shifted ({\it solid
contour}). The three components are separated by $\pm$1.5 km s$^{-1}$
from the systemic velocity (see Table 4). Different methods for
separating the outflow and quiescent components were tested but these
did not change the results within the overall uncertainties. The
integrated intensity maps of CO 4--3 are shown in
Fig. \ref{4:fig:CO4-3map}.  These are comparable to the CO 3--2 maps, but not with the HCO$^+$ 4--3 maps.

The HCO$^+$ 4--3 and CO 3--2 maps for Ced 110 IRS 4 and CrA IRAS 32
(see Fig. \ref{4:fig:HH100map}) show spatially resolved cores,
centered on the known IR position. For both sources, outflow emission
is seen originating at the source position.  HH 100
(Fig. \ref{4:fig:HH100map}) shows no HCO$^+$ 4--3 core, but does have
outflowing gas in the CO 3--2 map, originating at the IR position
(marked with a X).  There is strong HCO$^+$ emission north of HH100, which does not correspond to any source in the \citet{Nutter05} SCUBA maps. At this moment, this discrepancy cannot be explained. The maps around RCrA IRS 7A and 7B
(Fig. \ref{4:fig:HH46map}, sources marked with X and centered on RCrA
IRS 7A) show a single elongated core in the HCO$^+$ map that is up to
60$''$ along the major axis.  Both sources lie firmly within the
core. The $^{12}$CO 3--2 outflow is centered on
RCrA IRS 7A, suggesting it is the driving source.
HH 46 (Fig. \ref{4:fig:HH46map}) shows a nearly circular core in
 HCO$^+$, but a strongly outflow-dominated emission profile in CO 3--2,
with much stronger red-shifted than blue-shifted emission. For more
information, see \citet{vanKempen09a}.
The CO 4--3 integrated intensity maps
show elongated spatial profiles for Ced 110 IRS 4, Ced 110 IRS 6 and
Cha IRN, but no peak for Cha INa 2. IRAS 12496-7650 shows a round
core, dominated by outflow emission (see $\S$ 4.3.3)

\placeFigureChapterFourIRAS
\placeFigureChapterFourIRASspec

\subsection{JCMT map of IRAS 15398-3359}

Figure \ref{4:fig:int15398} shows the spectrally integrated intensity
maps of CO 3--2, HCO$^+$ 4--3 and C$^{18}$O 3--2 around the position
of IRAS 15398-3359. The maps clearly show a single core around the
central position, which is completely spatially resolved. There is
little to no C$^{18}$O or HCO$^+$ emission outside of the core, while
the CO 3--2 has emission lines at around 10 K in the surrounding
cloud. The C$^{18}$O map, which lacks outflow emission, is somewhat
shifted to the south from the other two maps. The spectra at the
central source position taken from these maps can be found in figure
\ref{4:fig:spec15398}. The CO spectra are dominated by the
blue-shifted outflow. The integrated intensity of HCO$^+$ 4--3 can be
found in Table \ref{4:tab:res2ahco}.
The independently calibrated CO 3--2 spectra
taken with JCMT HARP-B differ from the spectrum taken with APEX-2a by
only 3$\%$ in integrated intensity, well within the stated calibration
uncertainties of both telescopes.

\section{Analysis}
\subsection{Envelope properties}

Assuming the (sub-)mm emission is dominated by the cold dust in the
protostellar envelope, its mass can be derived from dust continuum
observations \citep{Shirley00, Jorgensen02}. Using the SCUBA Legacy
archive \citep{diFrancesco08}, envelope masses were derived for YSOs
in Corona Australis. \citet{Henning93} observed
fluxes at 1.3 mm for a large sample in Chamaeleon using the SEST,
which are used here to constrain the masses of the very southern
sources.  All masses were
derived using the relation
\begin{equation}
M_{tot}=S_\nu D^2 /\kappa_\nu B_\nu (T_{\rm{d}})
 \end{equation}
 assuming an isothermal sphere of 20 K, a gas-to-dust ratio of 100 and
 a dust emissivity $\kappa_\nu$ at 850 $\mu$m of 0.02 cm$^2$
 g$^{-1}$(gas+dust) from \citet[column 5,][]{Ossenkopf94}.  To scale
 to 1.3 mm, the dust emissivity was assumed to scale with frequency as
 $\kappa_\nu \propto \nu^{1.5}$ following the results from \citet{Ossenkopf94}.  The mass of the HH 46 envelope is
 estimated at $\sim$5 M$_\odot$ using new LABOCA observations and
 radiative transfer modelling \citep{vanKempen09a}. For IRAS
 12496-7650, LABOCA data of Cha II was used (Nefs, van Kempen \& van
 Dishoeck, in prep) The radii $R_{\rm{core}}$ over which $S_\nu$ is
 calculated are taken to be the core radii from
 \citet{diFrancesco08}. Such radii are not the FWHM of the cores, but
 are quite close to the radius where the temperature reaches 10~K
 \citep{Jorgensen02}.  No mass could be derived for Cha IRS 6a, RCrA
 TS 3.5, IRAS 07178-4429, Cha INa2 and IRAS 13546-3941 as they were
 not detected or observed.  The estimated envelope masses are given in
 Table \ref{4:tab:column}. They range from 0.04 to 6.3 M$_\odot$. For
 Corona Australis, our masses are a factor of three lower than those
 by \citet{Nutter05}. Differences are due to the difference in assumed
 dust temperature of 12 vs 20 K, as well as a different dust
 emissivity $\kappa_\nu$ at 850 $\mu$m.

 With the assumption that the H$^{13}$CO$^+$ 4--3 and C$^{18}$O 3--2
 emission is optically thin, one can estimate the column densities of
 these molecules in the protostellar envelope. In this calculation, an
 excitation temperature of 20 K is assumed, based on the average
 temperature within the protostellar envelope, for both the envelope
 and cloud material. Table \ref{4:tab:column} gives the
 column densities of H$_2$, C$^{18}$O and H$^{13}$CO$^+$, where the
 H$_2$ column density is derived from the C$^{18}$O column density
 assuming a H$_2$/CO ratio of $10^4$ and a $^{16}$O/$^{18}$O ratio of
 550 \citep{Wilson94}.  The H$_2$ column densities range from a few
 times 10$^{22}$ cm$^{-2}$ in Chamaeleon to a few times 10$^{23}$
 cm$^{-2}$ in Corona Australis, all in a 18$''$ beam. However,
 freeze-out of CO onto the interstellar grains make the adopted ratio
 of H$_2$ over CO a lower limit. In very cold regions ($T<$ 15 K),
 abundances can be as much as two orders of magnitude lower than the
 assumed CO/H$_2$ abundance of 10$^{-4}$. 
For sources where continuum data at 850 $\mu$m are available from
 \citet{diFrancesco08}, the column density was independently
 calculated from the dust, assuming a dust temperature of 20 K. These
 numbers are in general a factor of 2 to 3 higher consistent with some
 freeze-out of CO, but for some sources the column densities derived
 from the dust approach those obtained from C$^{18}$O.

The isotopologue observations of C$^{18}$O and H$^{13}$CO$^+$ also
allow to estimate the average optical depth $\tau$ in the CO 3--2
and HCO$^+$ 4--3 lines (Table \ref{4:tab:column}) following the formula of 
\begin{equation}
I(CO)/I(C^{18}O) = (1-e^{-\tau})/(1-e^{-\tau/550}))
\end{equation}
with the isotopologue ratios for both C$^{18}$O and H$^{13}$CO$^+$ taken from \citet{Wilson94}. For I, the peak temperatures from Tables 3 and 4 were taken to avoid contributions from outflow material. The optical depths range from 25 to a few hundred in the CO 3--2 and from about 0.9 to 16 for the HCO$^+$ 4--3. HCO$^+$ 4--3 is still quite optically thick, even though it is associated with just the envelope. One error in the estimates from the optical depths, may be that the C$^{18}$O 3--2 and H$^{13}$CO$^+$ themselves are not optically thin, but have optical depths of a few. In that case, the optical depth of the main isotope would lower as well. However, observations of even rarer isotopologues, such as C$^{17}$O or HC$^{18}$O$^+$ would be needed. 


Recent C$^{18}$O 3--2 data of the Ophiuchus cloud show that at many locations
throughout L~1688, multiple foreground layers are present providing
additional reddening of embedded sources
\citep{vanKempen09}. 
In our sample, all C$^{18}$O 3--2 lines can be fitted with
single Gaussians indicating that no fore-ground layers are present in any of the observed clouds, in
contrast with Ophiuchus.


\placeTableChapterFourSix

\subsection{Embedded or not?}

Analysis of the physical structure of several sources classified as
Class I based on their IR spectral slope has shown that some of them
are in fact not embedded sources, but rather edge-on disks, such as
CRBR 2422.8 \citep{Brandner00,Pontoppidan05}, or IRS 46
\citep{Lahuis06}.  Others turn out to be obscured (background) sources
or occasionally even disks in front of a dense core
\citep{vanKempen09}.  Theoretical models from \citet{Robitaille06} and
\citet{Crapsi08} indeed show that edge-on disks can masquerade as
embedded sources in their infrared SED due to the increased reddening.

Following \citet{Robitaille06} and \citet{Crapsi08}, the border
between a truly embedded so-called `Stage 1' source and a `Stage 2'
star + disk system is put at accretion rates of 10$^{-6}$ M$_\odot$
yr$^{-1}$, which in the context of these models corresponds to an
envelope mass of $\sim$0.1 M$_\odot$ or a H$_2$ column of $(2-6)\times
10^{21}$ cm$^{-2}$.  Because of potential confusion with surrounding
(foreground) clouds and disks, one cannot simply use the masses
derived from the continuum data but one must look in more detail at
the properties of a source. One of the key additional criteria for
identification as a truly embedded source is the concentration of warm
dense gas around the source. The concentration is determined by the
comparison of the peak intensity in a map with the amount of extended
emission.  Such a parameter has been extensively discussed and
successfully applied in the analysis of submillimeter continuum
observations from SCUBA \citep{Johnstone01,
Walawender05,Jorgensen07}. For example, they have shown that
pre-stellar cores have much more extended and less concentrated
emission than protostellar envelopes.

A recent survey of Class I sources in the Ophiuchus L~1688 region has
demonstrated that such a method can be expanded to molecular line
mapping and be used as a powerful tool for identifying truly embedded
protostars from edge-on disks and obscured sources
\citep{vanKempen09}. HCO$^+$ 4--3, a high density tracer, proved to be
particularly useful. The recipe for proper identification consists of
three ingredients. First the peak intensity of HCO$^+$ line should be
brighter than 0.4 K, a factor of 4 more than is commonly seen for
disks \citep{Thi04}. Second, the concentration parameter of HCO$^+$
4--3 should be higher than 0.6. Third, the HCO$^+$ should preferably
be extended within the single dish beam. For non-resolved HCO$^+$
emission, other information, such as whether or not the source is
extended in the submillimeter continuum or whether it has a (compact)
outflow, should be taken into account.

 The concentration of HCO$^+$ is defined as:

\begin{center}
\begin{equation}
C_{\rm{HCO^+}}=1-\frac{1.13 B^2 S_{\rm{HCO^+}}}{\pi R^2_{\rm{obs}} T_{\rm{int}}},
\end{equation}
\end{center}
with $S_{\rm{HCO^+}}$ the HCO$^+$ emission integrated spatially (K km s$^{-1}$) and 
spectrally over the entire envelope with radius $R_{\rm{obs}}$ (in arcseconds), and
$T_{\rm{int}}=\int T_{\rm MB} dV$ the HCO$^+$ emission integrated
spectrally over the central beam $B$ of 18$''$ alone (in K km s$^{-1}$.  The observed
core radius is taken to be the FWHM of the spatial distribution of the
integrated HCO$^+$ intensity (in arcseconds).  These values of $R_{\rm{obs}}$,
tabulated in Table \ref{4:tab:concen}, are often comparable to the
core radii found from the continuum data.  Although less reliable due
to a lower critical density and potential contributions from outflows,
the quiescent part of the CO 4--3 emission, which is associated with
the central regions of protostellar envelopes, can serve as a
diagnostic of the concentration of the warm dense gas in the inner
envelope as well. Note that any outflow wings have to be removed.

Table \ref{4:tab:concen} lists the derived concentration parameters
for our sources. Of the 12 sources for which such data exist, 6 have
concentrations of $\sim$0.7 and higher, as well as strong HCO$^+$
fluxes above 0.5 K km s$^{-1}$ associated with the sources,
characteristic of embedded sources.  Three sources (HH
100, IRAS 07178-4429 and IRAS 13546-3941) have no associated HCO$^+$
core and thus no concentration. 
Similarly, Cha INa2 shows no signs of centrally concentrated CO 4--3
emission. 
Cha IRS 6a, Cha INa2, RCrA TS 3.5, IRAS 07178-4429 and IRAS 13546-3941
all lack HCO$^+$ 4--3 emission, even below the limit adopted from
\citet{Thi04}, and are thus very likely not embedded YSOs.

HH 100 is an exception.  Although no concentration could be calculated
for this source as it is not peaking on the HH 100 source position, there is strong extended HCO$^+$ emission with at least 7$\sigma$, and in some areas up to 50$\sigma$. A
clear outflow is seen in the CO emission in Fig. \ref{4:fig:HH100map},
thus indicating that HH 100 is an embedded source.  This is confirmed from results from Spitzer-IRAC and MIPS (Peterson, priv. comm.). However, the lack of concentration from an envelope is puzzling due to the low optical depth derived for this line at HH100-OFF.

The RCrA IRS 7 binary system was classified as two embedded sources
by \citet{Groppi07}. The elongated HCO$^+$ spatial emission profile
indeed suggests that two envelopes are located close to each
other. The concentrations of 0.68 and 0.66 (close to 0.7) together
with the strong outflow confirm that both sources are
embedded. Similar arguments are used to classify RCrA IRS5 and IRAS
12496-7650 as embedded YSOs. In summary, using our line diagnostics,
we have confirmed that eleven of our sixteen Class I sources are truly
embedded YSOs, and five sources very likely not.


\placeTableChapterFourSeven 
\placeTableChapterFourEight

\subsection{Outflows}
\subsubsection{Temperatures}
 If we assume that these line wings fill the beam and are tracing the dominant part of the outflowing gas mass (also known as the swept up gas), then the ratios of emission in the line wings of the CO 3--2, 4--3 and 7--6
profiles at the central position in Table \ref {4:tab:outflow} 
can be compared to model line ratios from Appendix C of \citet{vanderTak07},
derived using statistical equilibrium calculations within an escape
probability formalism. The line wings are assumed to be optically thin
and filling factors to be the same within the different beams.  A quick look at the line wings of CO 3--2 and C$^{18}$O 3--2 in the sources in Corona Australis provides an upper limit to the optical depth of the CO line wing of about 3.1 close to the quiescent part of the line. At more extreme velocities, the opacity is probably much lower and indeed optically thin. Optically thin line wings were also shown to be a good assumption for the case of HH 46 by
\citet{vanKempen09a}.  Kinetic temperatures can then be derived if
outflow emission is detected in two of these three lines. Note that an
observed ratio leads to a degenerate solution of the kinetic
temperature and local density and not a unique solution of the
temperature.  Thus, some estimate of the local density must be
available.  Densities in the outer envelope can be estimated from
radiative transfer modelling of the dust continuum emission and SED
\citep{Ivezic97}. H$_2$ densities of a few times 10$^5$ cm$^{-3}$ are
derived by \citet{Jorgensen02} for distances of 1500--3000 AU
(corresponding to 10-20$''$ at the distances of Chamaeleon and Corona
Australis), but these drop to a few times 10$^4$ cm$^{-3}$ at a few
thousand AU. These densities are assumed to be comparable to the
densities in the swept-up gas at these distances.

At densities above the critical density,
$n_{\rm{cr}}$=5$\times10^4$ cm$^{-3}$ for CO 3--2, the gas is
thermally excited and a temperature can be inferred directly. For
lower densities, the gas is subthermally excited and higher kinetic
temperatures are needed to produce similar ratios of the line wings.
To estimate the temperatures, two scenarios are presented.  In the
first, it is assumed that the gas is thermally excited ($n$ $>$
$n_{\rm{cr}}$) since densities in the envelope are near to, or higher
than, the CO 3--2 critical density. This is applicable for all sources
with the exception of HH~46, for which the central beam extends to
about 4,500 AU in radius. The kinetic temperatures determined in this
way are lower limits.  A second scenario is considered by assuming
that the densities are of order 10$^4$ cm$^{-3}$, closer to typical
densities of the surrounding cloud or envelopes at larger radii.  In
the first scenario, outflow temperatures are on the order of 50
K.  An exception is IRAS 15398-3359, which appears to be much
warmer (100 K) from both the CO 3--2/4--3 ratio as well as the
4--3/7--6 ratio in the blue outflow. Since we lack CO 4--3 and/or 7--6
data of the red outflow, no temperature could be determined for the
red outflow. If densities are lower (scenario 2), derived kinetic
temperatures are on the order of 150 K, although for HH~46 (blue) and
Ced 110 IRS 6 (both red and blue) the outflow temperatures are
consistent with $\sim$80 K. Errors on the outflow temperatures are generally on the order of 25 to 40 $\%$, due to the sensitive dependency on density, as well as errors on the estimate  of the CO line ratios. 

Bow-shock driven shell models by \citet{Hatchell99} predict
temperatures of the swept-up gas by calculating the energy balance
between the heating due to the kinetic energy of the expanding
momentum-conserving shell and the line cooling of the gas. For
parameters typical of the Class 0 L~483 source ($L$=9 L$_\odot$,
$M_{\rm{env}}$= 4.4 M$_\odot$), outflow temperatures of 50 to 100 K
are found in the inner few thousand AU from the star. Along the bulk
of the outflow axis, temperatures are modelled to be 100--150 K.
Higher temperatures in excess of a few hundred to a thousand K are
predicted only near the bow shocks. The models depend on the local
conditions such as dust cooling, density distribution and jet
velocity, but most parameters have an influence of the order of a few
tens of K on the temperature along most of the outflow.  The results
in Table \ref{4:tab:outflow} agree well with the temperature
predictions from \citet{Hatchell99} and suggest that jet velocities
$>$100 km s$^{-1}$ are not present, with the exception of IRAS
15398-3359.

Although we assumed the line wings to be optically thin, this is not
necessarily the case \citep[e.g.,][]{Cabrit92,Hogerheijde98}. A more
in-depth discussion of the effects of optical depth and density on the
kinetic temperatures of the swept-up gas in the molecular outflow of
HH 46 IRS is presented in \citet{vanKempen09a}. 
A lower line ratio for the same temperature can also be caused by a higher optical depth in one of the line wings.
 Similarly a higher
density will lead to lower inferred temperatures for the same observed
line ratio. \citet{Hatchell99} also show that there may be a temperature gradient away from the connecting line between bow shock and origin at the protostar. However, this gradient is often diluted out and may be observable within a single beam. The different temperatures of two different ratios may be responsible for this.

\placeFigureChapterFourTwelve

\subsubsection{Other outflow parameters}
For the six sources where outflow emission has been mapped in
$^{12}$CO 3--2, it is possible to determine the spatial outflow
parameters, such as outflow mass, maximum velocity and
extent. Subsequently the dynamical time scales and average mass
outflow rate, the outflow force (or momentum flux) and kinetic
luminosity can be derived following a recipe detailed in
\citet{Hogerheijde98} with the assumption that the emission at the
more extreme velocities is optically thin and that the material has 
kinetic temperatures of 50-80 K.  For IRAS 15398-3359 a
temperature of 100 K is adopted. The inclination of the outflow with
the plane of the sky influences the results and the derived values
need to be corrected for their inclination using an average of the
model predictions in \citet{Cabrit90}. For HH 46 the inclination has
been constrained to 35$^\circ$ \citep{Reipurth91,Micono98}. For the
other sources, an inclination is estimated from the distance between
the major and minor axes of the maximum contour levels in
Figs. \ref{4:fig:HH100map} and \ref{4:fig:HH46map} (see Table
\ref{4:tab:outflow1}).

The results in Table \ref{4:tab:outflow1} show that most outflows have
masses of 0.01-0.1 M$_\odot$, with average mass loss rates between
2$\times10^{-6}$ to 4.4$\times10^{-4}$ M$_\odot$ yr$^{-1}$, and
outflow forces $F_{\rm{CO}}$ of 6.4$\times10^{-6}$ to
2.7$\times10^{-2}$ M$_\odot$ yr$^{-1}$ km s$^{-1}$. The strongest
outflow is RCrA IRS 7, which produces the highest outflow force and
the most kinetic luminosity. This is mainly caused by the fact that emission
is detected out to $\Delta V_{\rm{max}}$ as high as 23 km s$^{-1}$, compared
with 4 to 9 km s$^{-1}$ for the other sources, almost an order of
magnitude difference. Note that for HH 100 (red only), HH 46 (red
only) and RCrA IRS 7 (red and blue), the quoted outflow radius is a
lower limit, due to insufficient coverage. The resulting outflow force
and kinetic luminosity are thus upper limits. For HH 100 and and RCrA
IRS 7 subtracting Gaussians with widths of 3 km s$^{-1}$ changes the
integrated intensity by 20$\%$. All other parameters depend linearly
on the mass $M$ that is derived from the integrated intensity.

\citet{Hogerheijde98} surveyed the outflows of Class I sources in the
Taurus cloud. There, most sources produce weaker outflows than found
in Table \ref{4:tab:outflow1}, with outflow forces of the order of a
few times 10$^{-6}$ M$_\odot$ yr$^{-1}$ km s$^{-1}$, with a few sources having
values as high a few times 10$^{-4}$ M$_\odot$ yr$^{-1}$ km s$^{-1}$.
A recent survey of the outflows present in the Perseus cloud shows
$F_{\rm{CO}}$ values of a few times 10$^{-7}$ M$_\odot$ yr$^{-1}$ km
s$^{-1}$ to a few times 10$^{-5}$ M$_\odot$ yr$^{-1}$ km s$^{-1}$
\citep{Hatchell07a}. However, these numbers were not corrected for
inclination, which can account for at least a factor of 5
increase. The observed outflow forces in our sample are all among the
upper range of these values, especially HH 100 and RCrA IRS 7. Figure
5 of \citet{Hatchell07a} shows that most Class I outflows have lower
outflow forces than the Class 0 sources.

Most flows observed from the northern hemisphere were first
discussed in \citet{Bontemps96} using CO 2--1 observations with
relatively large beams. Outflow forces range from a few times
10$^{-7}$ to 10$^{-4}$ M$_\odot$ yr$^{-1}$ km s$^{-1}$, corrected for
inclination and optical depth. In \citet{Bontemps96}, a relation between the
outflow forces, bolometric luminosities and envelope masses was
empirically derived,
\begin{math}
\log(F_{\rm{CO}}) = -5.6 + 0.9 \log (L_{\rm{bol}})
\end{math}
and
\begin{math}
\log(F_{\rm{CO}}) = -4.15 + 1.1 \log (M_{\rm{env}}).
\end{math}

Figure \ref{4:fig:forces} shows a comparison between the observed
outflows of Table \ref{4:tab:outflow1} and the relations from
\citet{Bontemps96}.  Although the outflows
in our sample are all associated with Class I sources, they are
exceptionally strong outflows for their luminosity, but not
for their envelope mass. Outflow forces are one to two orders of magnitude
higher than expected from the above equation with luminosity. 
This suggests that the outflow
strength as measured through the swept-up material says more about the
surroundings (i.e., the amount of matter that can be swept up) than
about the intrinsic source and outflow properties.

\placeFigureChapterFourEleven

\subsubsection{The outflow of IRAS 12496-7650}

An outflow was identified from the line profiles in
\citet{vanKempen06} and it was concluded that the emission as seen
with ISO-LWS \citep{Giannini01} has to originate on scales larger than
8$''$ and not from the inner regions of a protostellar envelope around
IRAS 12496-7650.  Figure \ref{4:fig:I12496} reveals that the detected
blue-shifted outflow is located only in the central 15$''$ and is
peaking strongly on the source.  The lack of red-shifted emission reinforces the conclusion
that the outflowing gas is viewed almost perfectly face-on and that
our line of sight is straight down the outflow cone.

\subsubsection{The outflow of IRAS 15398-3359}

The line profiles of IRAS 15398-3359 have a clear blue outflow
component, with red-shifted emission notseen until further off source.
If only the line wings are mapped, Figure \ref{4:fig:I15398} shows
that the outflow of IRAS 15398-3359 is small and extends only
$\sim$25$''$ across the sky.  As discussed above, the blue part of the outflow is exceptionally
warm, with significant outflow emission seen in the CO 7--6
line. Although it is smaller in extent, the outflow does not stand out
w.r.t to the others in terms of outflow properties as measured from
their CO 3--2 maps.

\placeFigureChapterFourIRASutflow

\section{Conclusions}
We present observations of CO, HCO$^+$ and their isotopologues,
ranging in excitation from CO 3--2 to CO 7--6 of a sample of southern
embedded sources to probe the molecular gas content in both the
protostellar envelopes and molecular outflows in preparation for
future ALMA and Herschel surveys. The main conclusions are the
following:
\begin{itemize}
\item HCO$^+$ 4--3 and CO 4--3 integrated intensities and
concentrations, combined with information on the presence of outflows,
confirm the presence of warm dense quiescent gas associated with 11 of
our 16 sources.

\item RCrA TS 3.5, Cha IRS 6a, Cha INa2, IRAS 07178-4429 and IRAS
  13546-3941 are likely not embedded YSOs due to the lack of HCO$^+$
  4--3 emission and/or the lack of central concentration in HCO$^+$
  4--3.


\item The swept-up outflow gas has temperatures of the order of
  50--100 K, as measured from the ratios of the CO line profile wings,
  with the values depending on the adopted ambient densities. These
  values are comparable to the temperatures predicted by the heating
  model of \citet{Hatchell99}, with the exception of IRAS 15398-3359,
  which may be unusually warm.

\item The outflows of 6 of our truly embedded sources --- Ced 110 IRS
  4, CrA IRAS 32, RCrA IRS 7A, HH 100, HH 46,IRAS 15398-3359 --- were
  characterized using CO spectral maps. The outflows all have
  exceptionally strong outflow forces, almost two orders of magnitude
  higher than expected from their luminosities following the relation
  of \citet{Bontemps96}.

\item Neither Chamaeleon nor Corona Australis have foreground layers
  as found in Ophiuchus L~1688. All C$^{18}$O 3--2 spectra can be
  fitted with single gaussians.


\end{itemize}

Future observations using ALMA and Herschel will be able to probe the
molecular emission associated with these YSOs at higher spatial
resolution and higher frequencies. Comparison with single dish data,
both continuum surveys and spectral line mapping, will be essential in
analyzing the results obtained with future facilities.\\


\begin{acknowledgements}

 TvK and astrochemistry at Leiden Observatory are supported by a
Spinoza prize and by NWO grant 614.041.004. The APEX staff, in
particular Michael Dumke, are thanked for their extensive support and
carrying out the observations.  We are also grateful for the
constructive comments of the anonymous referee.

\end{acknowledgements}
\bibliographystyle{../../../bibtex/aa}
\bibliography{../../../biblio}

\end{document}